\documentclass[linenumbers,twocolumn]{aastex631}

\newcommand\s{$\sigma$}
\nolinenumbers
\newcommand\sd{$-$}
\usepackage{multirow}
\newcommand{\R}{\textcolor{red}}
\newcommand{\B}{\textbf}

\shorttitle{JWST Transmission Spectrum of Sub-Neptune TOI-732~c}

\shortauthors{Rigby et al.}

\begin{document}

\title{A JWST Transmission Spectrum of the Temperate Sub-Neptune TOI-732~c}

\author[0009-0001-9140-3299]{Frances E. Rigby} \affiliation{Institute of Astronomy, University of Cambridge, Madingley Road, Cambridge CB3 0HA, UK}

\author[0000-0002-4869-000X]{Nikku Madhusudhan} \affiliation{Institute of Astronomy, University of Cambridge, Madingley Road, Cambridge CB3 0HA, UK}

\author[0000-0002-2705-5402]{Subhajit Sarkar} \affiliation{School of Physics and Astronomy, Cardiff University, The Parade, Cardiff CF24 3AA, UK}

\author[0009-0008-9026-5534]{Lorenzo Pica-Ciamarra} \affiliation{Institute of Astronomy, University of Cambridge, Madingley Road, Cambridge CB3 0HA, UK}

\author[0000-0002-0931-735X]{M\aa ns Holmberg} \affiliation{Space Telescope Science Institute, 3700 San Martin Drive, Baltimore, MD 21218, USA}

\author[0000-0002-8837-0035]{Julianne I. Moses}
\affiliation{Space Science Institute, Boulder, CO 80301, USA}

\correspondingauthor{Nikku Madhusudhan}
\email{nmadhu@ast.cam.ac.uk}

\begin{abstract}
\nolinenumbers
In recent years, JWST has facilitated detections of carbon-bearing molecules in the atmospheres of temperate sub-Neptunes orbiting M dwarfs, ushering in a new era in the characterization of this intriguing planetary regime. We report the transmission spectrum of the temperate sub-Neptune TOI-732~c, observed with JWST NIRISS, NIRSpec G395H and MIRI LRS between 0.9-12 $\mu$m. The observations provide evidence for methane (CH$_4$) in a H$_2$-rich atmosphere, at a volume mixing ratio of $\sim$1\%, and non-detection of NH$_3$ and HCN, along with nominal constraints on other prominent molecules H$_2$O, CO and CO$_2$, which are typically expected in H$_2$-rich atmospheres. We conduct a comprehensive survey of 250 chemical species and find moderate to strong evidence (up to $\ln B\sim 5.9$, $3.9\sigma$) for additional absorption due to one or more complex molecules including higher-order hydrocarbons and/or sulfur-bearing molecules. The spectral features are strongly degenerate among these molecules and with methane, which we find at $\ln B=3.2-8.8$ (up to $3.0-4.6$$\sigma$) significance. Two complex molecules are preferred with at least moderate evidence ($\ln B \gtrsim 2.5$) in both the near- and mid-infrared, while several others show such evidence in at least one of the two wavelength ranges. The preferred molecules are found in trace quantities on Earth, with no significant sources identified in other planetary atmospheres, requiring future work to assess their physical plausibility in this planet. Future observations are required to resolve the degeneracies and place more robust constraints on these species. We highlight the need for further theoretical and experimental work to robustly characterize the atmospheric and internal composition of TOI-732~c and similar sub-Neptunes.

\end{abstract}


\section{Introduction} \label{sec:intro}

Detection surveys over the past decade have established that sub-Neptune planets dominate the exoplanetary demographics \citep[e.g.][]{fressin2013,Fulton2018}. With radii between those of Earth and Neptune, $\sim$1-4 R$_\oplus$, this regime spans planetary sizes not encountered in the solar system. The increasing numbers of sub-Neptunes orbiting nearby bright stars found by more recent surveys are making it feasible to observe their planetary atmospheres using transit spectroscopy. Sub-Neptunes are more conducive to such observations than Earth-like planets due to their larger size and possible low mean molecular weight atmospheres. JWST transit spectroscopy provides a promising avenue to study temperate sub-Neptunes and begin to place robust constraints on their nature. Crucially, constraints on the atmospheric composition from spectroscopic observations are required to resolve degeneracies in the possible internal structures of these planets. Generally, their bulk properties can be explained by a wide range of internal structures, spanning mini-Neptunes, water worlds including hycean worlds, and gas dwarfs \citep[e.g.][]{sotin2007,rogers2010a,valencia2013,Howe2014,Madhusudhan2020,Nixon2021,Rigby2024,Rigby2024b}.

Molecular detections have now been made in the atmospheres of two temperate sub-Neptunes with JWST \citep[e.g.][]{Madhusudhan2023b,Holmberg2024,Benneke2024}. The presence of multiple spectral features for key molecules in the near-infrared means that JWST has been able to provide precise abundance constraints on a number of such molecules. Near-infrared observations of the habitable zone sub-Neptune K2-18~b revealed detections of CO$_2$ and CH$_4$ in its H$_2$-rich atmosphere, with an absence of NH$_3$ and H$_2$O \citep{Madhusudhan2023b}. A similar pattern of detections was found for TOI-270~d, in addition to tentative inferences of H$_2$O and CS$_2$ \citep{Holmberg2024,Benneke2024}. Thanks to considerations on photochemistry and possible surface-atmosphere interactions, abundance constraints for molecular species in the photosphere can provide a window into the planetary surface and interior, helping to break internal structural degeneracies. For instance, the abundances of key molecular species including CH$_4$, CO$_2$, CO, NH$_3$ and H$_2$O can indicate the presence or lack of a deep atmosphere \citep[e.g.][]{Tsai2021,Yu2021,Hu2021,Madhusudhan2023a,Rigby2024b}. A deep atmosphere scenario would be consistent with either a gas dwarf -- a rocky interior beneath a thick H$_2$-rich envelope -- or mini-Neptune interior.

Recent mid-infrared observations of K2-18~b using JWST MIRI revealed possible evidence for other molecules. Based on an exploration of 20 relevant molecules, \citet{Madhu2025} reported moderate evidence for dimethyl sulfide (DMS) and/or dimethyl disulfide (DMDS) \citep{Madhu2025}, both predicted to be biomarkers \citep{domagal-goldman2011, seager2013b}, with DMS additionally suggested from the near-infrared observations \citep{Madhusudhan2023b}. Considering an expanded set of 90 hydrocarbons, \citet{Welbanks2025} found comparable evidence to DMS/DMDS for six other molecules using one of the MIRI datasets reported in \citet{Madhu2025}. Considering an even larger set of 650 chemical species, and both the NIR and MIR data, \citet{PicaCiamarra2025} found three molecules with at least moderate evidence in the MIR and potential signs in the NIR. These molecules are DMS, diethyl sulfide, and methyl acrylonitrile \citep{PicaCiamarra2025} -- of these only DMS has been predicted in observable quantities in these atmospheres \citep[e.g.][]{seager2013b,Tsai2024}. Most recently, \citet{Luque2025} conducted a joint analysis of the NIR and MIR datasets with a limited set of molecules and claimed no significant evidence for DMS/DMDS using the original data from \citet{Madhusudhan2023b} and \citet{Madhu2025}. This is consistent with the findings of \citet{Madhusudhan2023b} when considering the same retrieval set-up as \citet{Luque2025} with two detector offsets in the NIR data. However, using their own data reduction of the same data sets, \citet{Luque2025} find moderate evidence for DMS (Bayes factor of 2.3-2.8, or 2.7-2.9$\sigma$). Overall, while none of these studies report strong evidence ($>$3.6$\sigma$) required to claim a robust detection, they nevertheless present some evidence for excess absorption due to DMS and/or other complex molecule(s), motivating further investigation. Additional observations of K2-18~b are required for a more robust inference of DMS, DMDS, or other potential molecules.

In this work, we present the first JWST transmission spectrum of the temperate sub-Neptune TOI-732~c (also known as LTT~3780~c). TOI-732~c orbits its M3.4V host star at $0.08$~au \citep{Cloutier2020,Nowak2020}, with equilibrium temperature $305$~K ($353$~K) at albedo $A_\mathrm{B}=0.5$ ($A_\mathrm{B}=0$). The planet, discovered with TESS, has been observed by multiple studies \citep{Cloutier2020,Nowak2020,Bonfanti2024}, resulting in three mass and radius measurements. The most recent measurement by \citet{Bonfanti2024} is reported to be $M_\mathrm{p}=8.04^{+0.50}_{-0.48}\ \mathrm{M_\oplus}$ and $R_\mathrm{p}=2.39^{+0.10}_{-0.11}\ \mathrm{R_\oplus}$, with this $M_\mathrm{p}$ intermediate between the previous measurements. We adopt these most recent values from \citet{Bonfanti2024} in this study. Based on the planetary bulk properties and the presence of a H$_2$-rich atmosphere \citep{Cabot2024}, the density of TOI-732~c could be explained by a range of interior compositions, including a gas dwarf, mini-Neptune, or hycean scenario \citep[e.g.][]{Cloutier2020,Rigby2024}. The TOI-732 system also contains the rocky super-Earth TOI-732~b, with $M_\mathrm{p}=2.46\pm0.19\ \mathrm{M_\oplus}$, $R_\mathrm{p}=1.325^{+0.057}_{-0.058}\ \mathrm{R_\oplus}$, and $T_\mathrm{eq}\sim900$~K \citep{Bonfanti2024}. The two planets lie either side of the radius valley -- the observed paucity of planets with radii $\sim$1.5-2.0 $\mathrm{R}_\oplus$ \citep[e.g.][]{Fulton2017,Fulton2018,Cloutier2020b}. Systems such as TOI-732 provide an interesting testing ground for hypothesis on the mechanism leading to the radius valley around M dwarf hosts \citep{Cloutier2020,Cloutier2020b,Bonfanti2024}.

The system is of additional interest due to the potential habitability of TOI-732~c, as the planet was proposed as a hycean candidate \citep{Madhusudhan2021}. Atmospheric observations are key for breaking degeneracies in possible internal structures of TOI-732~c and indeed similar such sub-Neptunes, which can help to identify their nature, and potential to host a liquid water ocean. Recently, \citet{Cabot2024} analyzed ground-based high-resolution transmission spectroscopic observations of TOI-732~c. They report marginal evidence for CH$_4$ at 2.2$\sigma$, and no evidence for the presence of NH$_3$. The presence of high-altitude clouds is also inferred through the muted spectral features at short wavelengths, equivalent to the JWST NIRISS coverage \citep{Cabot2024,Cheverall2024}.

With $T_\mathrm{eq}=305$~K at $A_\mathrm{B}=0.5$, TOI-732~c has equilibrium temperature intermediate between fellow hycean candidates K2-18~b and TOI-270~d, at $250$~K and $326$~K respectively (for $A_\mathrm{B}=0.5$). TOI-732~c is more observationally favourable than K2-18~b -- its host star is smaller and brighter, and the planet has lower surface gravity and a higher equilibrium temperature. The star has also been shown to exhibit low activity levels \citep{Sairam2025}.

In this work, we report the first JWST transmission spectrum of this planet, TOI-732~c. The spectrum was obtained using NIRSpec, NIRISS and MIRI instruments covering the wavelength range 0.9-12 $\mu$m. We derive constraints on the atmospheric composition, which provide the first insight into the possible nature of the planet. In Section \ref{sec:obs}, we present the JWST observations, followed by atmospheric retrievals in Section \ref{sec:retrieval}, which we conduct for both the near-infrared (NIR) and MIRI datasets. Finally, our results and directions for future work are discussed in Section \ref{sec:discussion}. 

\section{Observations and Data Reduction} 
\label{sec:obs}

We report the first transmission spectrum of TOI-732~c with JWST using  NIRISS \citep{Doyon2012,Doyon2023}, NIRSpec \citep{Ferruit2012,Birkmann2014} and MIRI \citep{Rieke_2015}.  Three transit observations were performed as part of GO Program 3557 (PI: N. Madhusudhan). The first observation was made with NIRISS with the Single Object Slitless Spectroscopy (SOSS) mode, on April 20th 2024 from 19:04:33 UTC to 23:37:32 UTC, leading to a total exposure time of 4.67 hours. The GR700XD grism ($R\sim$700) was used in combination with the CLEAR filter.  The configuration used the SUBSTRIP256 subarray and the NISRAPID readout pattern. The main exposure consisted of 632 integrations with 2 groups per integration.  An additional F277W exposure was included, consisting of 10 integrations, with 12 groups (up-the-ramp) per integration.  Target acquisition was performed on the main target.

The second transit was observed with NIRSpec, using the G395H grating with the F290LP filter ($R\sim$2700) in the Bright Object Time Series (BOTS) mode. The observation was conducted on May 27th 2024, from 13:43:12 UTC to 17:57:13 UTC, with a total exposure time of 4.60 hours. The configuration utilized the SUB2048 subarray and the NRSRAPID readout pattern. The full dispersed spectrum fell over two detectors, NRS1 and NRS2, spanning 2.73-5.17 $\mu$m, with a gap between the detectors at 3.72-3.82 $\mu$m.  The exposure consisted of 1543 integrations with 7 groups (up-the-ramp) per integration. Target acquisition was obtained using the nearby object 2MASS10183398-1143258 since the main target would cause detector saturation.

The third transit observation was conducted using the MIRI Low Resolution Spectrometer (LRS) \citep{Kendrew2015, Bouwman2023} with the SLITLESSPRISM subarray and the P750L prism.  The observation was performed on December 9th 2024, from 15:15:39.115 UTC to 18:51:39.762 UTC, with a total exposure time of 3.6 hours.  The FASTR1 readout mode was used.  There was a total of 5821 integrations with 13 groups per integration.  Using the JWST Exposure Time Calculator, we estimated a maximum fraction of saturation of 48\%.  An ancillary background observation was also performed, consisting of 30 integrations with 13 groups per integration. Target acquisition was obtained on the main source. No high-gain antenna movements were recorded during any of the observations.

We follow the same approach to data reduction as performed for K2-18 b \citep{Madhusudhan2023b, Madhu2025} and TOI 270 d \citep{Holmberg2024}.  NIRSpec and NIRISS data are reduced using the \texttt{JExoRES} pipeline.  For MIRI, given the higher uncertainties in this instrument, we apply both the \texttt{JExoRES} and \texttt{JexoPipe} pipelines for greater robustness.

\begin{figure}
	\includegraphics[width=0.45\textwidth]{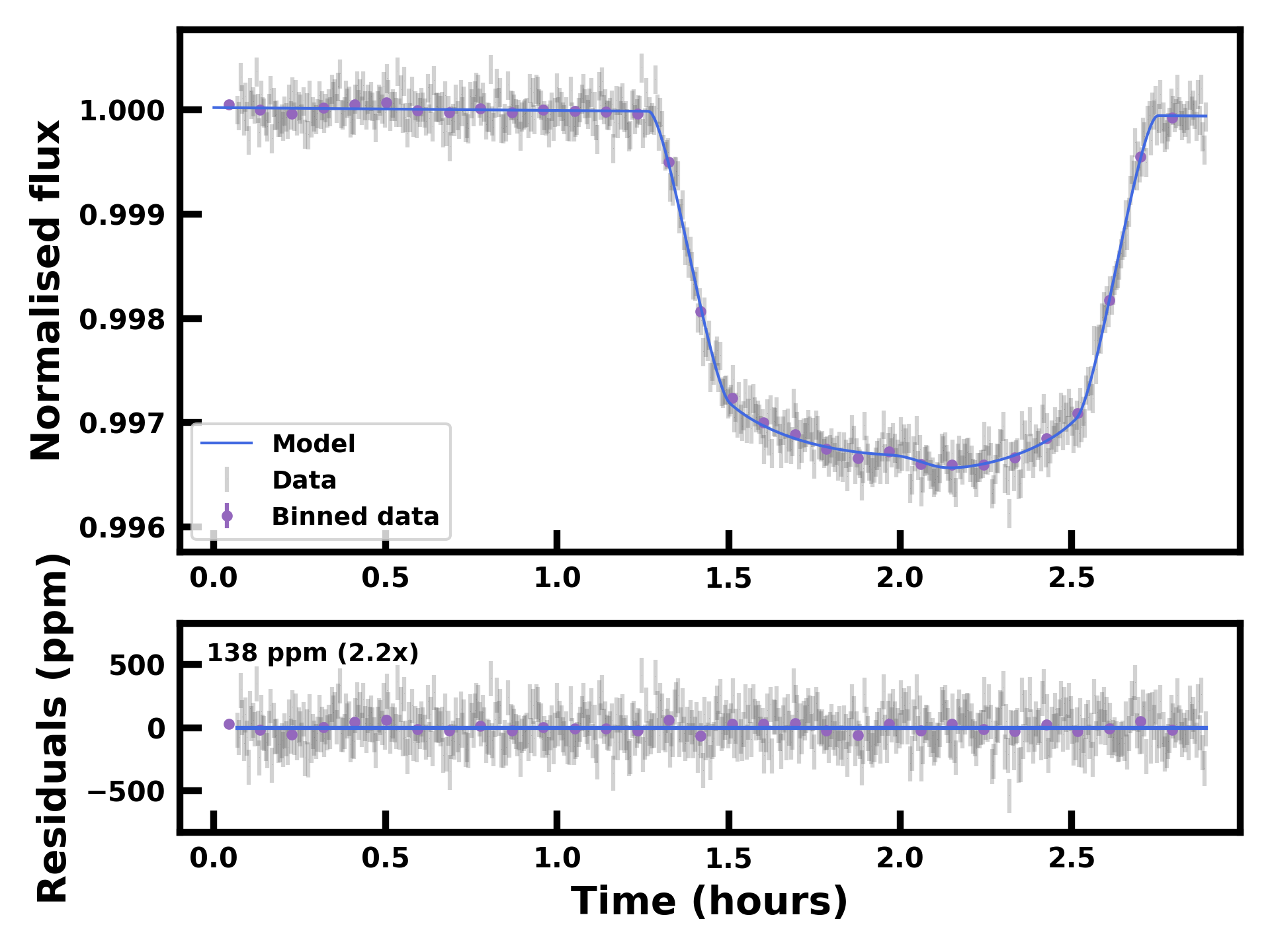}
    \includegraphics[width=0.45\textwidth]{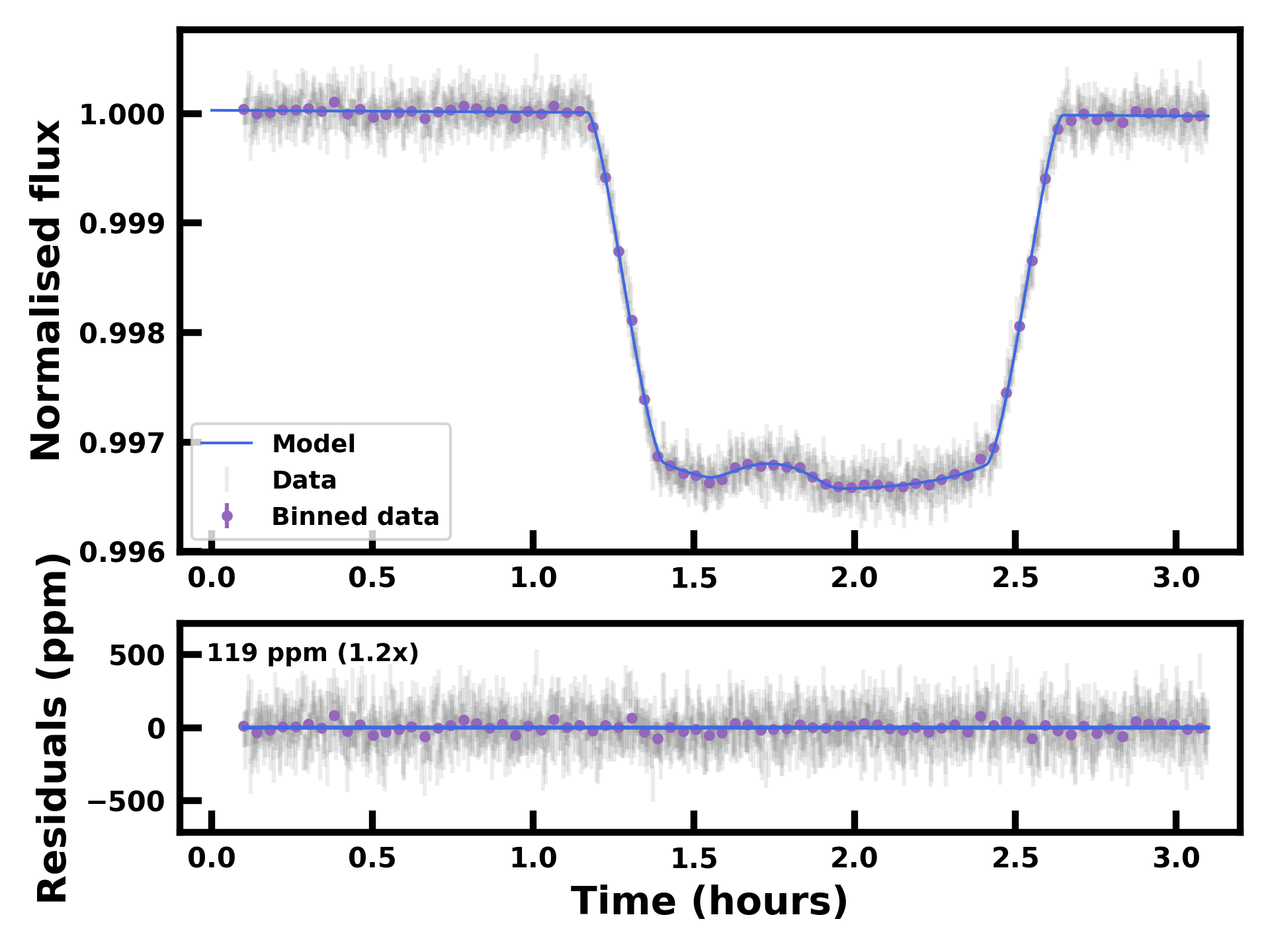}
     \includegraphics[width=0.45\textwidth]{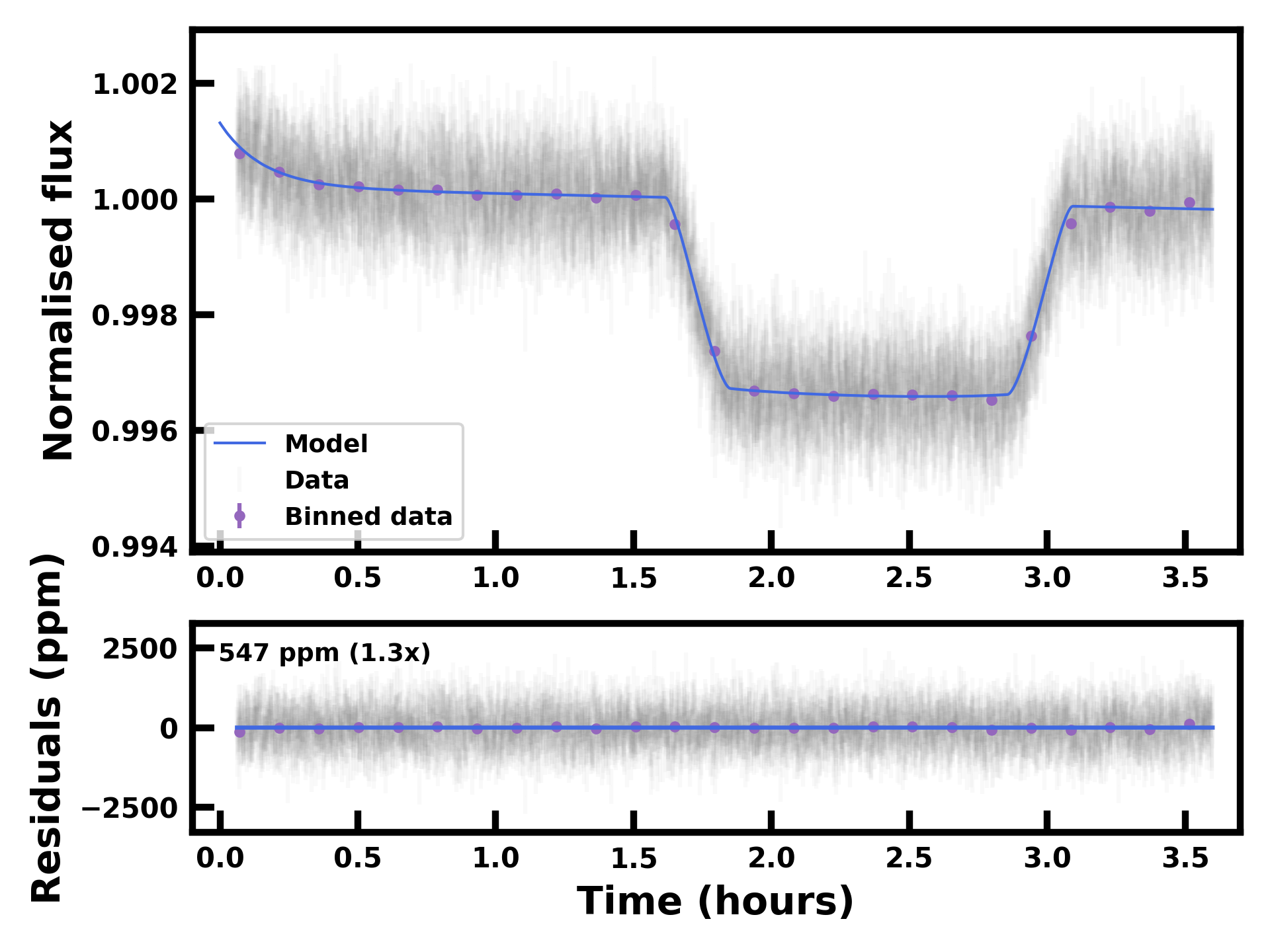}
    \caption{White light curves for the primary transit of TOI-732~c obtained with the three different instruments: NIRISS SOSS (top), NIRSpec G395H (middle) and MIRI LRS (bottom). In each case, the white light curve obtained using the \texttt{JExoRES} pipeline is shown along with the residuals from subtracting a median model fit. For NIRSpec, the combined white light curve from the two detectors NRS1 and NRS2 is shown, while for NIRISS we show the white light curve for Order 1. The standard deviation of the residuals for NIRISS, NIRSpec and are MIRI are 138 ppm, 119 ppm and 547 ppm, respectively.  These are $\sim$2.2, $\sim$1.2, and $\sim$1.3 times the expected level, respectively. The white light curves for NIRISS and NIRSpec show planet crossing of stellar heterogeneities during transit, as discussed in section~\ref{sec:obs}.}

    \label{fig:wlc_nirspec} 
\end{figure}

\subsection{NIRSpec}

We used a version of \texttt{JExoRES} to process the data \citep{holmberg2023}. Stage 1 of the pipeline begins with the raw .uncal files, and applies a combination of official JWST Science Calibration Pipeline \citep{Bushouse2020} steps and custom steps.  We include superbias subtraction, saturation flagging, reference pixel correction, linearity correction, dark current subtraction and jump detection (5$\sigma$ threshold). A group-wise background subtraction step is applied prior to ramp fitting to remove both 1/f noise and sky background. This involves obtaining the mean in each pixel column for pixels that are $\pm10$ pixels away from the middle of the spectral trace, after masking for bad pixels and cosmic-ray hits. The mean is then subtracted from all pixels in that column. Stage 1 ends with linear ramp fitting and the production of .rateints files with ramp images for each integration.  

In stage 2, the 2-D ramp images are further processed, including the assignment of the wavelength solution, and identification of outliers. Flat fielding is omitted. Stage 2 ends with the production of .calints files containing the time-series of 2-D spectral images.  In stage 3, the 1-D stellar spectra per integration are extracted using an optimal extraction algorithm \citep{horne_optimal_1986}, after first applying a bad pixel mask based on data quality flags.  We discard spectral channels with more than 20\% of the flux masked from further analysis.

\subsection{NIRISS}

We use the latest version of the \texttt{JExoRES} pipeline to analyse the NIRISS data set, as pursued in previous works \citep{holmberg2023, Madhusudhan2023b}. Stage 1 starts with the raw .uncal files and applies similar steps to those in the NIRSpec pipeline, including saturation flagging, superbias subtraction, linearity correction, and fitting of the group-level ramp. We skipped the jump step due to the low number of groups. We add a custom 1/f noise subtraction step just before the linearity correction. This involves subtracting off a model of the stellar (modulated by the white light curve) and background signals to reveal the 1/f noise on each group image. We then subtract this from the corresponding column in the corresponding main science image. The initial model used for the stellar and background signal is a group-wise median image.

In stage 2, a flat field correction is applied followed by a custom sky background subtraction step \citep{Madhusudhan2023b}.  The latter is a two step process (due to the known discontinuity in the background count at around pixel column 700).  Firstly, we scale the standard NIRISS zodiacal light background model (originally created from program 1541) for pixels columns $>\sim$ 700 followed by subtraction of this scaled background from the main science images. The scaling is performed by obtaining the median of all integration images in a small rectangular background region. For pixel columns up to $\sim700$ we first obtain a median integration image, and then find the median background count per pixel column, after masking off the spectral trace and bad pixels.  These column-wise medians are then subtracted from all science images up to pixel column $\sim 700$. In Stage 3, we performed  outlier correction, order tracing \citep{holmberg2023}, wavelength calibration using PASTASOSS \citep{Baines2023a, Baines2023b}, and box extraction with an aperture of 35 pixels. We extract the first order spectrum for our final analysis. The pipeline is then iterated again with an updated white-light curve and background model for the 1/f step, until convergence.

As part of Stage 3, we also identified and corrected contamination from a dispersed field star. Similar to \cite{holmberg2023}, we identified this source using GAIA DR3 \citep{Gaia2023}, and found that the G$_\mathrm{BP}$-G$_\mathrm{RP}$ magnitude was similar to that of WASP-96. We then reduced the NIRISS WASP-96 observation from JWST program~2734, scaled the flux according to the relative brightness difference between the contamination star and WASP-96, and subtracted this WASP-96 image from our data. The initial guess for the spatial position of the contamination star was based on GAIA DR3, followed by cross-correlating to refine the position. Overall, we found that performing this contamination correction did not significantly impact the transmission spectrum of TOI~732~c.

\subsection{MIRI}

We employ the same approach to our MIRI data reduction as described in \cite{Madhu2025}, applying two independent pipelines, \texttt{JExoRES} and \texttt{JexoPipe}.  For \texttt{JExoRES} the pipeline is as follows.  In stage 1, starting with the .uncal files, we perform data quality initialization, electromagnetic interference (EMI) correction, saturation flagging, first and last frame flagging, linearity correction, reset switch charge decay (RSCD) correction, dark-current subtraction, and ramp fitting.  These are all steps from the JWST Science Cailbration pipeline.   In Stage 2, we perform flat-fielding.  We then use the gain reference file to convert units from DN/s to e$^-$/s, followed by outlier detection for cosmic rays.  This involves sigma-clipping on the time series of each pixel, using a rolling median and a threshold of $7\sigma$.  These pixels together with their neighbors are masked, as are any other pixels with bad data quality flags.  A background subtraction is then performed for each integration image, using two regions which are 6-23 pixels away from the center of the trace. The mean count per pixel row in the background region is obtained and then subtracted from all pixels in that row.  In Stage 3, we perform 1-D spectral extraction using an extraction aperture of 9 pixels, using optimal extraction.  Further sigma-clipping is performed on the light curves themselves.

For \texttt{JexoPipe}, we start with the raw .uncal files and in stage 1, perform the following steps from the JWST Science Cailbration pipeline:  group scale, data quality initialization, EMI correction, saturation flagging, first frame and last frame correction, reset correction, linearity correction, RSCD correction, dark subtraction, jump detection (threshold of $5\sigma$), ramp fitting and gain scale. In stage 2, we combine all stage 1
segments into a single .rateints file and then apply the assign World Coordinate System (WCS) and flat field steps.  This is followed by a custom bad-pixel flagging step \citep{Sarkar2024}, that flags pixels with abnormal DQ values and 3$\sigma$ outliers (found on a row-by-row basis) with NaN values. We then perform a custom background subtraction on each integration, taking pixel columns 10-14 and 60-66 to sample the background region.  After applying an outlier mask we obtain the mean in each row and then subtract this from all pixels in that row.  Next, we apply a custom bad pixel correction step that uses temporal and spatial interpolation to fill in all bad pixel locations. We then apply another step to correct for outliers using a rolling median (of 20
contiguous integrations).  Outliers $\pm 5 \sigma$ from the rolling median value are replaced by the median value. In Stage 3, the 1-D stellar spectra are extracted using an optimal extraction algorithm \citep{horne_optimal_1986}, after application of an aperture of 9 pixels width.
 
\subsection{Light curve fitting}
\label{light_curve_fitting}

We next proceed to light curve fitting.  We first take the 1-D stellar spectra time series and construct white light curves for each of the three observations.  We use these to obtain system parameters, and then proceed to fitting of spectral light curves.  When fitting for limb darkening we use the formulation and priors of \cite{Kipping2013}.  

\subsubsection{White light curves}
\label{WLC_fitting}

For NIRISS we obtain the white light curve from the first order, while for NIRSpec  we combine the NRS1 and NRS2 time series to obtain the white light curve.  Figure \ref{fig:wlc_nirspec} shows the white light curves from each observation.  Evident in the curves for NIRISS and NIRSpec are the apparent effects of occulted stellar heterogeneities. Therefore, for these two observations, we use the semi-analytical spot modeling code \texttt{SPOTROD} \citep{Beky2014} as described in \citep{Madhusudhan2023a} to obtain a joint  inference of transit parameters, systematic trend parameters and spot parameters.  Parameter estimation uses nested sampling with \texttt{MultiNest}.  We fit for the planet-to-star radius ratio ($R_p/R_s$), mid-transit time ($t_0$), the normalized
semi-major axis ($a/R_s$), the inclination angle ($i$), two quadratic limb darkening coefficients (LDCs) ($u_1$  and $u_2$), two parameters for a linear trend, and four spot/faculae parameters (size, contrast, and two coordinates). We assume a circular orbit with period of 12.252284 days \citep{Bonfanti2024}. In the likelihood, we also include a parameter to inflate the light curve error bars to match the residual scatter between the data and the transit light-curve model.  The fitted system parameters for each white light curve are given in Table \ref{tab:wlc_params}.

For NIRISS, we found that the white light curve may be explained be an occulted star spot or faculae, with some preference for a faculae. However, the necessary size of the spot/faculae is very high, approaching a radius of 0.5~R$_\star$, which we used as the upper limit for the prior. Because of the preference for a faculae, we adopted this explanation. However, the apparent size may be unrealistic. Future work can explore other alternatives for explaining the data. 

For MIRI, spot occultations are not apparent in the white light curve. Therefore we do not apply \texttt{SPOTROD}  and instead use analytic models for the transit. 
For \texttt{JExoRES}, we use the analytic transit model from \texttt{batman} \citep{kreidberg_batman_2015}, multipled by a systematic trend function.  The fit is performed using nested sampling with \texttt{Multinest}.  The trend function has exponential and linear components.  The overall model has the form: $F_{obs}(t) = F_{out} (1 + \alpha\tau + \gamma e^{- \tau/ \delta} ) F_{transit}(t)$, where $\tau$ is the time since the start of the observation, $F_{transit}$ is the \texttt{batman} transit model, and $F_{out}$, $\alpha$, $\gamma$, and $\delta$ are parameters in the trend function.   The first 200 integrations are also masked to remove the initial steep decline due to detector settling.  The white light curve is obtained from wavelengths between 4.8-12 $\mu$m.  
We then fit for $R_p/R_s$, $t_0$, $a/R_s$, $i$, $u_1$  and $u_2$, and the four parameters for the trend.  

In \texttt{JexoPipe}, we use wavelengths between 5-10 $\mu$m to construct the MIRI white light curve.  We again exclude the first 250 integrations.  Outliers on the white light curve $\pm 2.5\sigma$ from a rolling median are identified and the spectra on those integrations replaced by linear interpolation of spectra from adjacent integrations.
 We use a transit model from \texttt{PyLightCurve} \citep{Tsiaras2016} and the same same linear/exponential trend function as above.  The out-of-transit residuals are estimated by fitting the trend function only and the error bars on the light curve are scaled so that the average error bar equals the standard deviation of the residual scatter.
 We use \texttt{emcee}\citep{foreman-mackey_emcee_2013} to perform MCMC parameter estimation on the white light curve, fitting for $R_p/R_s$, $t_0$, $a/R_s$, $i$, $u_1$  and $u_2$, and the four parameters for the trend.  The fitted white light parameters are given in Table \ref{tab:wlc_params}.

\subsubsection{Spectral light curves}
\label{SLC_fitting}

For the NIR spectroscopic light curve fitting, we fix $t_0$ from the corresponding white light curve, and $a/R_s$ and $i$ to the values from NIRISS white light curve (which used the NIRSpec constraints as priors). The spot/faculae parameters are also fixed to the corresponding white light fit values with the exception of the spot/facula contrast (which is wavelength-dependent). Given the high impact parameter of TOI-732~c ($b > 0.7$), we adopt model limb darkening parameters from ExoCTK \citep{bourque2021} using an ATLAS9 stellar model \citep{Castelli2003}. For this, we adopted a bin width of 2 pixel per wavelength bin for NIRISS and NIRSpec. This way, we obtain a final final high resolution transmission spectrum for each observation.

For MIRI, given that in the mid-infrared we expect minimal wavelength-dependent variation in the LDCs, we fix the LDCs for spectral light curves to those obtained from the MIRI white light curve fit from the respective pipeline data.  We bin the light curves with a width of 0.2 $\mu$m or 5 pixels whichever has more pixels, giving a minimum bin width of 0.2 $\mu$m.  Other aspects of the light curve fitting for each pipeline mirror that of the respective white light curve.  We fix $t_0$, $a/R_s$ and $i$ to the values from the corresponding white light curve fits and fit for $R_p/R_s$ and the four trend parameters, per light curve. We also note that TOI-732~b begins eclipsing just before the egress of planet~c, and that explicitly fitting for both planets does not significantly alter the MIRI transmission spectrum of TOI-732~c.

\begin{figure*}
	\includegraphics[width=\textwidth]{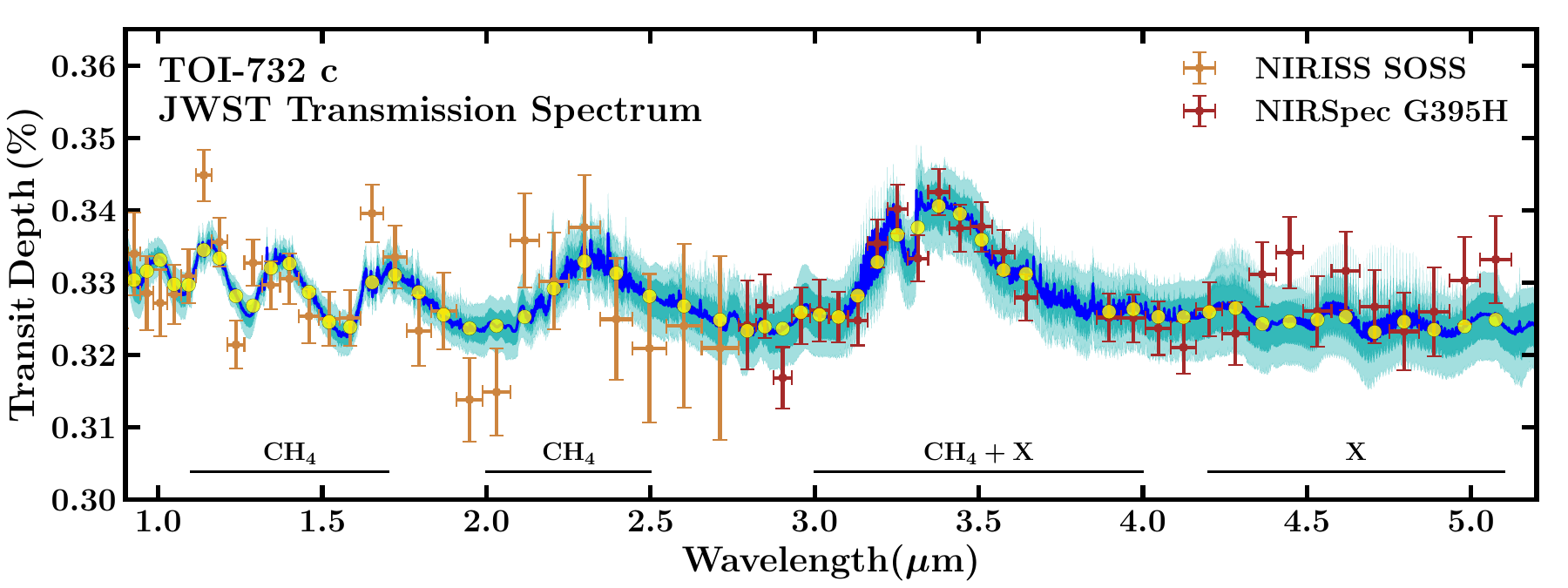}
    \caption{The transmission spectrum of TOI-732~c in the NIR. We show the spectrum and retrieved model fits for the one-offset retrieval, as discussed in Section~\ref{sec:finalretrieval}. This model includes the 6 key CNO molecules (CH$_4$, CO$_2$, CO, H$_2$O, NH$_3$ and HCN) and the 2 species found to exhibit moderate preference across our retrieval cases, referred to as X, as discussed in Section~\ref{sec:tracespecies}. The data in orange and red show the NIRISS and NIRSpec spectra respectively, covering 0.9-2.8 $\mu$m and 2.8-5.2 $\mu$m. The retrievals were carried out using the native resolution data. For visual clarity, these are shown binned to resolutions of $R\approx25$ and $R\approx55$ for the NIRISS and NIRSpec data respectively. The NIRSpec spectrum has been vertically offset by the median retrieved value for the one-offset retrieval, by $\approx 112$ ppm. In blue we show the median retrieved spectrum, with the lighter blue contours indicating the $1\sigma$ and $2\sigma$ intervals. The points in yellow show this median spectrum binned to the resolution of the observations.}
    \label{fig:spectrum} 
\end{figure*}

\begin{table*}
\centering
\begin{tabular}{lcccc}
\hline \hline
Parameter & NIRISS & NIRSpec & MIRI& MIRI \\ 
&  & &  (\texttt{JExoRES}) & (\texttt{JexoPipe})\\ \hline
Midtransit time, T$_0$  & $60420.941727_{-0.000060}^{+0.000064}$ &  $60457.698845_{-0.000024}^{+0.000022}$ & $60653.735400_{-0.000060}^{+0.000060}$ & $60653.735275_{-0.000067}^{+0.000070}$\\ 
(BJD - 2400000.5) & & & & \\
Inclination, $i$ ($^{\circ}$) & $88.8855_{-0.0090}^{+0.0087}$ & $88.874_{-0.014}^{+0.013}$ & $88.8730_{-0.0095}^{+0.0095}$ &$ 88.865 _{-0.037}^{+0.039}$\\ 
$a / R_*$ & $42.24_{-0.23}^{+0.23}$ & $42.02_{-0.36}^{+0.33}$  & $41.99_{-0.24}^{+0.24}$ & $41.78_{-0.88}^{+0.96}$ \\ 
 $R_\mathrm{p} / R_*$ & $0.05868_{-0.00072}^{+0.00097}$ & $0.05860_{-0.00083}^{+0.00040}$ & $0.05819_{-0.00026}^{+0.00032}$ & $0.05897_{-0.00035}^{+0.00046}$ \\ \hline

\end{tabular}
\caption{Parameter estimation resulting from the analysis of the white light curves for each of the observations of TOI-732~c with NIRISS SOSS (order 1), NIRSpec G395H (NRS1 + NRS2) and MIRI LRS (using two different pipelines).  For NIRSpec and NIRISS observations, we model spot/facula crossings as described in Section \ref{light_curve_fitting}. The orbital period is fixed at $P = 12.252284$ days \citep{Bonfanti2024}. We used the constrains of $i$ and $a / R_*$ from NIRSpec as priors when fitting the NIRISS and MIRI data.
$a / R_*$ is the normalised semi-major axis and $R_\mathrm{p} / R_*$ is the 
planet-to-star radius ratio.}
\label{tab:wlc_params}
\end{table*}

\section{Atmospheric Retrieval} \label{sec:retrieval}

In this section, we describe our retrieval analysis of the transmission spectrum of TOI-732~c. Our approach begins with a retrieval using the near-infrared (NIR) data from NIRISS SOSS and NIRSpec G396H, spanning 0.9-5.2 $\mu$m, as shown in Figure~\ref{fig:spectrum}. The set-up of this retrieval and the chemical species included follows a similar method to the analysis of other temperate sub-Neptunes in the NIR \citep{Madhusudhan2023b,Holmberg2024}. We first conduct retrievals including a number of prominent CNO molecules and other relevant molecules previously considered in retrievals of sub-Neptune atmospheres \citep[e.g.][]{Madhusudhan2023b,Holmberg2024, Benneke2024, Madhu2025}. Following this, we conduct sequential retrievals for a large number of chemical species, along with the baseline CNO molecules which are unconstrained or allowed at significant abundances by the initial retrieval, following the approach in recent studies \citet{Madhu2025, Welbanks2025,PicaCiamarra2025}. We identify the chemical species that meet an evidence threshold $\ln B \geq2.5\pm0.5$ ($\geq$$2.7\pm0.2\sigma$), where $B$ is the Bayes factor, before proceeding to conduct equivalent retrievals with the MIR LRS data, spanning 5-12 $\mu$m. The $\ln B = 2.5$ (2.7$\sigma$) threshold is chosen as it corresponds to ``moderate'' evidence in Jeffrey's scale \citep[e.g.][]{Trotta2008}, and for the purposes of this study it is lowered to $\ln B = 2.0$ (2.5$\sigma$) to allow for a 0.5 uncertainty in $\ln B$ for a 2.5$\sigma$ result, equivalent to $\sim$$0.2\sigma$ uncertainty \citep{PicaCiamarra2025}. 

The transmission spectrum probes the properties of the atmosphere at the day-night terminator, which we model as an H$_2$-dominated plane-parallel atmosphere with uniform chemical composition, in hydrostatic equilibrium. The pressure-temperature (P-T) profile, the chemical abundances, and the properties of clouds and hazes are all treated as free parameters in the retrieval framework. The P-T profile is modeled parametrically following \citet{Madhusudhan2009}. Inhomogeneous clouds and hazes are included in the model, with hazes above a grey cloud deck \citep{Pinhas2018}; the relevant parameters are the cloud-top pressure $P_c$, the haze Rayleigh enhancement factor $a$ and scattering slope $\gamma$, and the overall fraction covered at the terminator by clouds and hazes $\phi$. The atmospheric model and retrieval set-up we use has been outlined in previous studies \citep[e.g.][]{Madhusudhan2020, Constantinou2022,Madhusudhan2023b,Holmberg2024}. 

\subsection{Initial Retrieval} \label{sec:initialretrieval}
We first perform an atmospheric retrieval of the NIR data using the \texttt{AURA} retrieval code \citep{Pinhas2018} following the approach of similar previous studies of temperate sub-Neptunes K2-18~b and TOI-270~d \citep[e.g.][]{Madhusudhan2023b,Holmberg2024}. We include molecular opacity contributions from 6 CNO molecules typically considered to be prominent in the H$_2$-rich atmosphere of a temperate sub-Neptune, similar to previous works \citep[e.g.][]{Constantinou2022,Madhusudhan2023b,Holmberg2024, Benneke2024}: H$_2$O, CH$_4$, CO, CO$_2$, NH$_3$ and HCN. We also consider five molecules which have been suggested as potential biomarkers for rocky planets \citep{Segura2005, domagal-goldman2011, seager2013a, seager2013b, catling2018, schwieterman2018} and hycean worlds \citep{Madhusudhan2021}. These are (CH$_3$)$_2$S (dimethyl sulfide or DMS), CS$_2$, CH$_3$Cl, OCS and N$_2$O, as considered in previous studies of hycean candidates \citep{Madhusudhan2023b,Holmberg2024,Madhu2025}. 

Following recent works, the molecular absorption cross sections are obtained using line lists as follows: H$_2$O \citep{Polyansky2018}, CH$_4$ \citep{Hargreaves2020}, CO \citep{rothman2010, Li2015}, CO$_2$ \citep{HUANG2013, huang2017}, NH$_3$ \citep{Coles2019}, and HCN \citep{harris2006, barber2014}. For each molecule, pressure broadening due to H$_2$ is considered, as in \citet{gandhi2020}. Collision-induced absorption is also included for H$_2$-H$_2$ and H$_2$-He \citep{borysow1988,orton2007,abel2011,richard2012}. Line lists from the \textsc{HITRAN} database \citep{HITRAN2016} were used to compute the absorption cross sections for CH$_3$Cl \citep{ch3cl_1, ch3cl_2}, OCS \citep{ocs_1, ocs_2, ocs_3, ocs_4, ocs_5, ocs_6, ocs_7} and N$_2$O \citep{n2o_2}. Following \cite{Madhusudhan2021}, the \textsc{HITRAN} \citep{dms_cs2_1,dms_cs2_2,HITRAN2016,HITRAN2020} cross sections at 1~bar and 298~K are used for the remaining molecules, DMS and CS$_2$. 

The \texttt{MultiNest} implementation of Nested sampling \citep{Feroz2009} via \texttt{PyMultiNest} \citep{Buchner2014} is used to perform the Bayesian inference and parameter estimation. The model parameter priors are outlined in full in Appendix~\ref{sec:priors}. As described above, in our initial retrieval we include 11 chemical species, hence the retrieval includes 11 free parameters corresponding to their mixing ratios. There are an additional 6 free parameters to describe the $P$-$T$ profile \citep{Madhusudhan2009}, 4 for clouds/hazes, and 1 for the reference pressure, giving 22 free parameters in total for this model. We define the reference pressure $P_\mathrm{ref}$ as the pressure at the median planetary radius $R_\mathrm{p}=2.39\ \mathrm{R_\oplus}$ \citep{Bonfanti2024}.

Previous studies have allowed for parametric offsets between the spectra obtained from the NIRISS and NIRSpec instruments as well as between the NRS1 and NRS2 detectors used in NIRSpec G395H \citep[e.g.][]{Madhusudhan2023b, moran_high_2023}. Here, similarly to \citet{Madhusudhan2023b}, we consider two offset scenarios: (1) one offset for the NIRSpec spectrum with respect to the NIRISS spectrum, and (2) separate offsets for each of the NRS1 and NRS2 NIRSpec spectra with respect to the NIRISS spectrum.

We first start with the same set-up as \citet{Madhusudhan2023b} for the NIRISS and NIRSpec spectra of K2-18~b, including the 11 molecules described above using the AURA retrieval code. We find strong contribution for CH$_4$, with multiple absorption features across the NIR wavelength range. We infer the log volume mixing ratio to be $\log(X_{\mathrm{CH}_4}$) = $-1.54^{+0.42}_{-0.71}$ for the one-offset retrieval. For consistency, we also conduct the same retrieval with the POSEIDON code and obtain similar results, as shown in Section~\ref{sec:res:molabundances}. Furthermore, the one-offset case is preferred over the two-offset case at $\ln B\sim2.0$ ($\sim$ 2.5$\sigma$). In all these cases, CH$_4$ is detected with a strong Bayesian evidence of over 4$\sigma$ relative to the model without CH$_4$. Among the other prominent CNO molecules, CO$_2$, CO, and H$_2$O have abundances that are relatively unconstrained by the retrieval, allowing for potentially high abundances.  We find strong upper limits on the volume mixing ratios for NH$_3$ and HCN, at $\log(X_{\mathrm{NH}_3})<-4.6$ and $\log(X_{\mathrm{HCN}})<-3.6$. The non-detection of NH$_3$ and the evidence for CH$_4$ are consistent with initial inferences using ground-based high-resolution spectroscopy of TOI-732~c \citep{Cabot2024}, as well as with similar chemical inferences for other temperate sub-Neptunes K2-18~b \citep{Madhusudhan2023b} and TOI-270~d \citep{Holmberg2024}. We also find tentative evidence for DMS, at $\ln B\sim2.0$ ($\sim$$2.3\sigma$), in the one-offset case for this model set-up. However, considering the possible degeneracies between DMS and other absorbers in the NIR, we explore potential contributions from other molecules as pursued in previous works \citep[e.g.][]{Tsai2021,Madhusudhan2023a, Madhu2025}.

We then consider additional molecules, which include C$_2$H$_4$, C$_2$H$_6$, CH$_3$OH, HC$_3$N, C$_5$H$_8$ (isoprene), C$_6$H$_6$, O$_2$, O$_3$, PH$_3$, SO$_2$, H$_2$S,  CH$_3$SSCH$_3$ (dimethyl disulfide or DMDS), and CH$_3$SH. In this set-up, there are a total of 24 molecules. For each molecule, we use the cross-sections from the \textsc{HITRAN} database \citep{dms_cs2_1,dms_cs2_2,HITRAN2016,HITRAN2020}, at 1~bar and 298~K where possible. Similarly to the \citet{Madhusudhan2023b} set-up with 11 molecules discussed above, we find strong evidence for CH$_4$, and constraints for CH$_4$, CO$_2$, CO, H$_2$O, NH$_3$ and HCN consistent with those noted above. In addition, we find that the spectral contribution from DMS is degenerate with those from a few other molecules, such as isoprene and DMDS. Therefore, the unique contribution from either of these molecules cannot be conclusively resolved. Overall, we find evidence for excess absorption beyond CH$_4$ in the NIR data due to more complex molecules. Therefore, to more robustly investigate the source of the excess absorption in this spectrum, we carry out an agnostic search to explore other molecules that could also explain it.

\begin{figure*}
	\includegraphics[width=\textwidth]{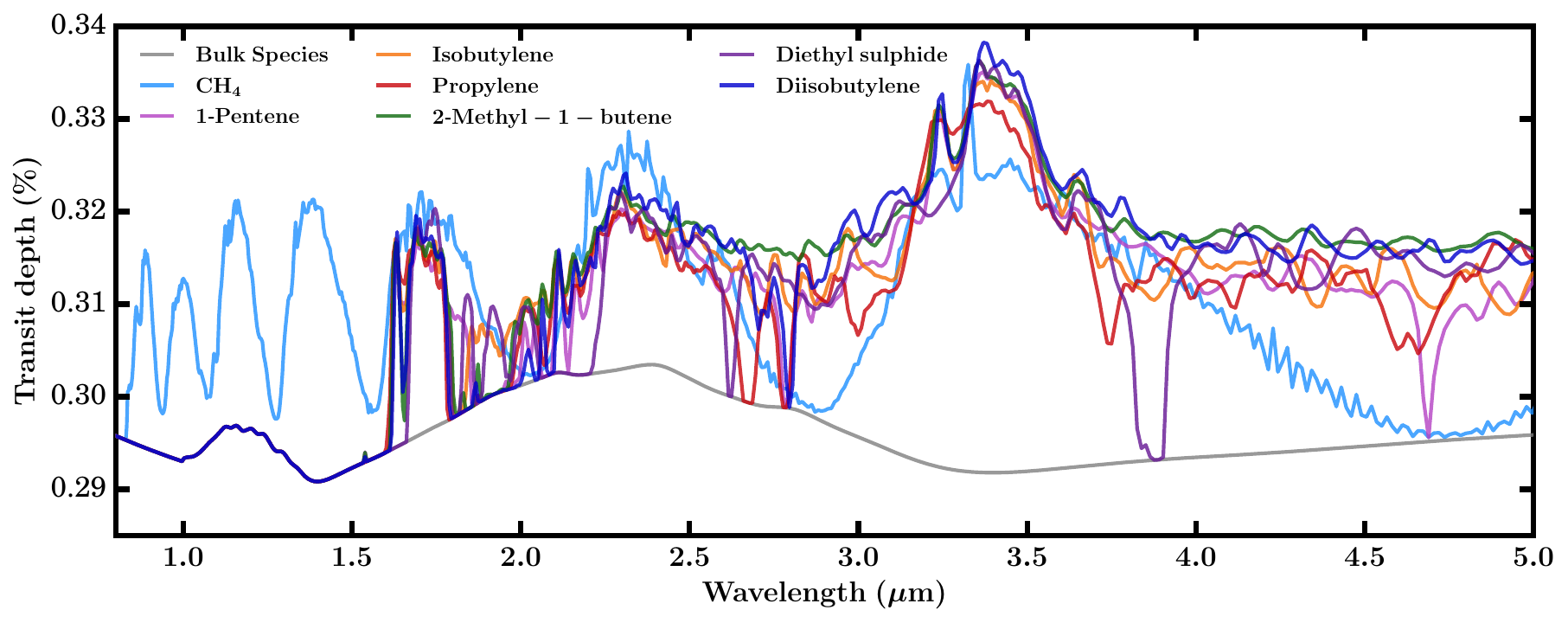}
    \includegraphics[width=\textwidth]{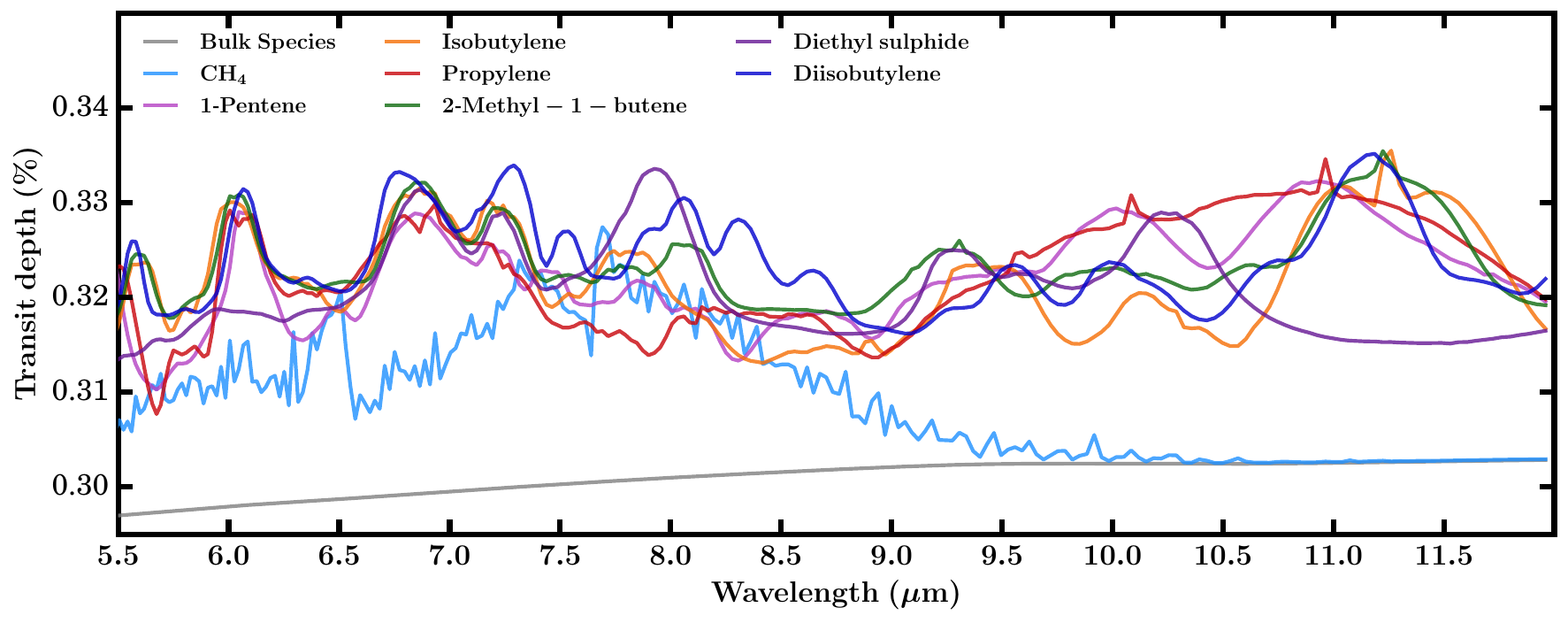}
    \caption{Contributions of several molecular species to the transmission spectrum in the 0.9-5.2 $\mu$m range (top) and 5-12 $\mu$m range (bottom). The individual contributions from each molecule is shown. This nominal model assumes an isothermal P-T profile at 350~K, and mixing ratios consistent with our retrievals: 10$^{-2}$ for CH$_4$, 10$^{-4}$ for all the other species. The contributions for each molecule shows the transmission spectrum from this molecule, in addition to collision-induced absorption of H$_2$-H$_2$ and H$_2$-He. The molecular species shown are those which exhibit moderate evidence across at least three of our four retrieval cases, as shown in Table~\ref{tab:poseidonevidence}.}
    \label{fig:contributions} 
\end{figure*}

\subsection{Search for Trace Species} \label{sec:tracespecies}

We now proceed to conduct a wide survey of possible chemical species, in order to thoroughly explore the potential atmospheric composition of TOI-732~c. We perform the sequential retrievals for the NIR and MIRI data following the approach in \citet{PicaCiamarra2025}, using the \texttt{POSEIDON} retrieval code \citep{macdonald2017,Macdonald2024}. In this section, we outline the retrieval procedure and set-up, before presenting the results for the NIR and MIRI retrievals.

We follow a similar model set up as with our \texttt{AURA} retrievals discussed above, except for the chemical species included. For K2-18~b, \citet{Madhu2025} consider 3-4 chemical species in each of their canonical retrievals for the MIRI data: CH$_4$, CO$_2$, and one or two additional chemical species X (DMS and/or DMDS). This was informed by the strong constraints placed on the abundances of CH$_4$ and CO$_2$, and non detections of other CNO species. As discussed in Section~\ref{sec:initialretrieval}, in the present case we place constraints on the CH$_4$ abundance in the NIR spectrum of TOI-732~c, and place upper limits on key molecules H$_2$O, CO$_2$, and CO, which are relatively unconstrained but have 95$\%$ upper limits allowing for high abundances. We find strong upper limits for NH$_3$ and HCN. Therefore, we include 5 chemical species in each of our present retrievals: CH$_4$, CO$_2$, H$_2$O, CO, and an additional chemical species X. 

We define this as the canonical model, i.e., a baseline of these 4 prominent species (CH$_4$, CO$_2$, H$_2$O and CO) combined with an additional species X, which we refer to as ``4+X''. The preference for X is established by comparing the Bayesian evidence of the canonical model to the baseline model; also see \citet{Madhu2025, Welbanks2025, PicaCiamarra2025}. The list of chemical species explored is presented in the Appendix and discussed below. With 5 chemical species this introduces 5 free parameters for their mixing ratios. There are additional 6 free parameters to describe the $P$-$T$ profile, 4 for clouds/hazes, 1 for the reference pressure, and 1 for an offset applied to the NIRSpec data relative to NIRISS, giving 17 free parameters in total for the models used in the \texttt{POSEIDON} retrievals. We again define the reference pressure $P_\mathrm{ref}$ at $R_\mathrm{p}=2.39\ \mathrm{R_\oplus}$ \citep{Bonfanti2024}. The model parameters and priors are listed in full in Appendix~\ref{sec:priors}.

We explore 250 chemical species as possible trace species X. As in \citet{PicaCiamarra2025}, we adopt cross-sections from the HITRAN database \citep{HITRAN2020} for $T\sim300$~K and $P\sim1$~bar, where these are available, and otherwise use the closest available to these values. The set of  chemical species from HITRAN span the following classes: hydrocarbons; nitriles, amines and other nitrogenated hydrocarbons; and sulfur-containing molecules. Additionally, we include cross-sections for six more species from the ExoMol database \citep{Tennyson2024} which are not available on HITRAN, and we compute the cross-sections for HC$_3$N using the line list available on HITRAN. With an additional 75 molecules already included in \texttt{POSEIDON}, this totals 250. The full list of molecules considered is given in Appendix~\ref{Appendix:Moleculeslist}.

As outlined above, we first carry out retrievals with this set-up for the combined NIRSpec G395H and NIRISS data, to identify molecules displaying evidence in this dataset. The threshold is set at $\ln B=2.0$ ($2.5\sigma$), following \citet{PicaCiamarra2025}, representing moderate evidence $\ln B=2.5\pm0.5$ (2.7 $\pm$ 0.2 $\sigma$), when allowing for a $\sim$0.5 uncertainty in computing the $\ln B$ (equivalent to $\sim$0.2$\sigma$). We first consider the case of one offset on NIRSpec relative to NIRISS, and then explore a case with two independent offsets for NIRSpec NRS1 and NRS2, both relative to NIRISS. We then proceed to carry out retrievals on the MIRI data, reduced via two independent pipelines, \texttt{JExoRES} and \texttt{JexoPipe}, to establish the set of molecules with moderate evidence in both wavelength ranges. In the following sections, we present our findings from these retrievals.

\subsubsection{Retrievals with NIR data}

In Table~\ref{tab:poseidonevidence}, we show the retrieved atmospheric abundances and model preferences for the chemical species we find with moderate evidence in the 4+X retrievals, exceeding $\ln B=2.5\pm0.5$ ($2.7\pm0.2\sigma$). In total, this is the case for 29 molecules. As in Section~\ref{sec:initialretrieval}, we consider two different combinations of possible offsets between the detectors: one offset between the NIRISS and NIRSpec spectra; and independent offsets for the NIRSpec NRS1 and NRS2 spectra. The number of molecules with moderate evidence reduces to 21 when considering two offsets. Introducing the two offsets somewhat reduces the $\ln B$ detection significance by $\sim$1 ($\sim0.1$-$0.2\sigma$) for the majority of molecules. These species are listed in Table \ref{tab:poseidonevidence} along with their respective model preferences. The spectral features of some of the prominent molecules are shown in Figure~\ref{fig:contributions}.

In Table~\ref{tab:poseidonevidence} we list the median retrieved log mixing ratio for each of these species. For the molecules with $\ln B < 2.0$ ($<2.5\sigma$) in the two-offset retrieval, we instead give the 95$\%$ upper limit. For the majority of these chemical species, the abundances are consistent between the one and two offset cases. Across the 29 molecules we typically infer median volume mixing ratios between $\sim$$10^{-5}-10^{-4}$ using the NIR data.

\subsubsection{Retrievals with MIRI data}

\begin{figure*}
	\includegraphics[width=\textwidth]{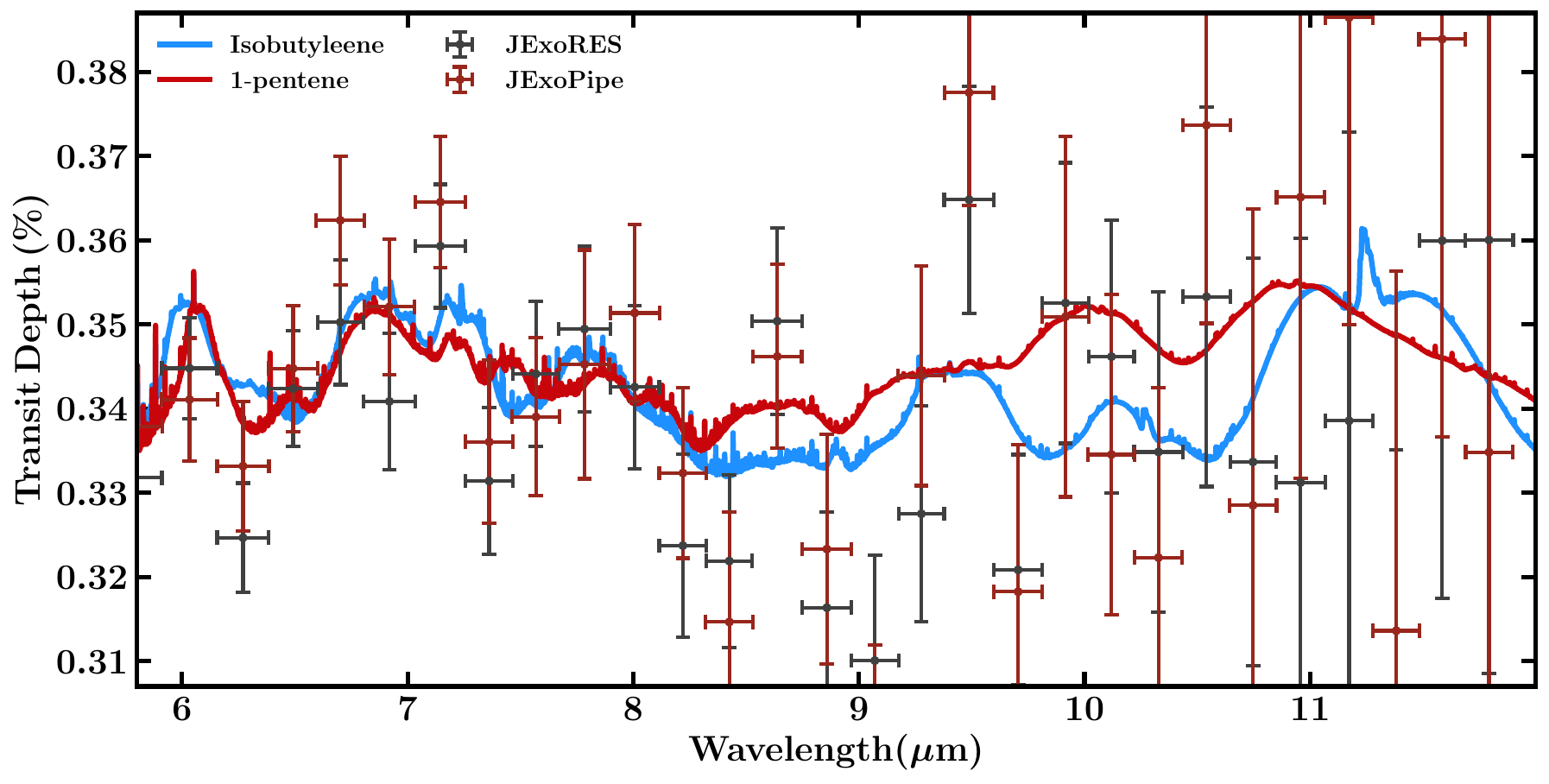}
    \caption{The MIRI \texttt{JexoPipe} and \texttt{JExoRES} data with retrieved spectral fits to the \texttt{JexoPipe} data. The spectral fits, represented by the solid curves, show the median retrieved spectra, obtained with \texttt{POSEIDON} retrievals on the MIRI \texttt{JexoPipe} data. In this set-up, the model includes the 4 baseline molecules (CH$_4$, CO$_2$, CO, H$_2$O) in addition to, in turn, either of the 2 molecules showing moderate evidence across all retrieval cases,  1-pentene and isobutylene.}
    \label{fig:spectrum_miri} 
\end{figure*}

For the 29 chemical species with moderate evidence ($\ln B\geq2.5\pm0.5$, or $\geq2.7\pm0.2\sigma$) in the NIR one-offset retrievals, we then proceed to carry out the equivalent retrieval using the MIRI data. We determine the model preferences for each of these species with respect to the baseline model including CH$_4$, CO$_2$, CO, and H$_2$O (4+X), as for the NIR retrievals. For robustness, we consider two independent reductions of the MIRI data with \texttt{JExoRES} and \texttt{JexoPipe}, similarly to \citet{Madhu2025} and \citet{PicaCiamarra2025}.

We find 8 molecules with moderate evidence in both the NIR and at least one MIR dataset. However, only 2 of these show at least moderate evidence across both MIRI reductions. These are isobutylene (also referred to as isobutene, C$_4$H$_8$) and 1-pentene (C$_5$H$_{10}$), as shown in Figure~\ref{fig:spectrum_miri}. For these 2 molecules, we infer log mixing ratios of $\sim$$10^{-4}-10^{-3}$ with the MIRI data, which are consistent with the NIR results to within the $1\sigma$ uncertainties, with generally lower inferred abundances from the NIR retrieval. The remaining 6 molecules, which reach moderate evidence in the NIR and the MIRI \texttt{JexoPipe} retrieval, are propylene (C$_3$H$_6$), 2-methyl-1-butene (C$_5$H$_{10}$), diisobutylene (diisobutene or 2,4,4-trimethyl-1-pentene, C$_8$H$_{16}$), DMS (CH$_3$SCH$_3$), isoprene (C$_5$H$_8$), and diethyl sulfide (C$_4$H$_{10}$S). We also note that a number of other molecules reach moderate evidence in the NIR only, including cyclopentene, which has the highest evidence of any molecule considered in the NIR, as shown in Table~\ref{tab:poseidonevidence}.

For K2-18~b, \citet{PicaCiamarra2025}, using a similar method to the present study, found only three molecules with potential evidence from the analysis of both the MIRI \citep{Madhu2025} and NIR \citep{Madhusudhan2023b} observations of K2-18~b. Those molecules were dimethyl sulfide (DMS), diethyl sulfide, and methylacrylonitrile.

As a term of comparison, we also consider a featureless (``flat line'') spectrum, and we calculate the Bayesian evidence for such a model. We consider two possible priors on the flat line's transit depth. In both cases, we centre the prior at the white light transit depth for the corresponding pipeline, and we allow for a range of $\pm 600$ ppm in one case, and $\pm 2000$ ppm in the other, following \citet{Madhu2025}. We find Bayesian evidences $\ln Z=$ 197.35-198.69 for \texttt{JexoPipe}, and $\ln Z =$ 205.58-206.61 for \texttt{JExoRES}, with the wider prior in both cases leading to the lower $\ln Z$. Compared to our best-fitting molecule, isobutylene, this results in Bayes factors $\ln B=$ 5.1-6.4 (3.6$\sigma$-4.0$\sigma$) for \texttt{JexoPipe} and $\ln B=$ 4.4-5.4 (3.4$\sigma$-3.7$\sigma$) for \texttt{JExoRES} against the flat line model. A comparison against our four-species baseline, with no excess absorption, instead results in significances between $\ln B= 1.0$ (2.0$\sigma$, \texttt{JexoPipe} with the narrow prior) and $\ln B = 2.84$ (2.9$\sigma$, \texttt{JExoRES} with the wide prior), still against the flat line model.

\begin{table*}
\centering

\begin{tabular}{l|cccc}
\hline \hline

Chemical species & \multicolumn{2}{c}{NIR} & \multicolumn{2}{c}{MIR} \\
& 1 offset & 2 offsets & \texttt{JExoRES} & \texttt{JexoPipe} \\
\hline
Cyclopentene & $-4.24^{+0.78}_{-0.79}$ (5.9, 3.9\s) & $-4.24^{+0.84}_{-0.81}$ (5.4, 3.7\s) & $<-1.23$ ($ -0.2 $, \sd) & $<-1.40$ (0.7, 1.8\s) \\
    \textbf{Isobutylene} & $\mathbf{-4.62^{+0.76}_{-0.58}}$ \textbf{(5.3, 3.7\s)}& $\mathbf{-4.50^{+0.74}_{-0.64}}$ \textbf{(4.5, 3.5\s)} & $\mathbf{-3.71^{+0.92}_{-1.47}}$ \textbf{(2.6, 2.8\s)} & $\mathbf{-3.78^{+0.85}_{-1.04}}$ \textbf{(4.1, 3.6}\s) \\
Propylene & $-3.94^{+0.80}_{-0.75}$ (5.3, 3.7\s) & $-3.96^{+0.84}_{-0.76}$ (4.5, 3.4\s) & $<-1.79$ (1.7, 2.4\s) & $-3.86^{+1.11}_{-1.20}$ (3.3, 3.0\s) \\
{\bf 1-pentene} & ${\bf -4.31^{+0.78}_{-0.74}}$ {\bf (4.7, 3.5\s)} & ${\bf -4.30^{+0.80}_{-0.77}}$ {\bf (3.8, 3.2\s)} & ${\bf -3.69^{+1.15}_{-1.33}}$ {\bf (2.1, 2.6\s)} & ${\bf -3.69^{+1.05}_{-1.29}}$ {\bf (2.9, 2.9\s)} \\
1-hexene & $-4.62^{+0.87}_{-0.71}$ (4.6, 3.5\s) & $-4.62^{+0.78}_{-0.69}$ (3.4, 3.1\s) & $<-1.64$ (1.1, 2.1\s) & $<-1.60$ (1.3, 2.2\s) \\
2-methyl-1-butene & $-4.21^{+0.83}_{-0.75}$ (4.5, 3.5\s) & $-4.17^{+0.86}_{-0.83}$ (4.3, 3.4\s) & $<-1.70$ (1.2, 2.2\s) & $-4.24^{+1.04}_{-1.22}$ (2.8, 2.9\s) \\
1-butene & $-4.11^{+0.82}_{-0.78}$ (4.3, 3.4\s) & $-4.11^{+0.83}_{-0.79}$ (3.8, 3.2\s) & $<-1.48$ (1.1, 2.1\s) & $<-1.65$ (1.6, 2.3\s) \\
2-methyl-1-pentene & $-4.35^{+0.71}_{-0.67}$ (4.2, 3.4\s) & $-4.35^{+0.75}_{-0.71}$ (3.6, 3.2\s) & $<-1.48$ (0.7, 1.8\s) & $<-1.75$ (1.7, 2.4\s) \\
1-heptene & $-4.83^{+0.75}_{-0.75}$ (4.0, 3.3\s) & $-4.80^{+0.79}_{-0.72}$ (2.9, 2.9\s) & $<-1.70$ (1.5, 2.3\s) & $<-1.86$ (1.6, 2.4\s) \\
4-methyl-1-pentene & $-4.40^{+0.83}_{-0.79}$ (3.7, 3.2\s) & $-4.56^{+0.84}_{-0.80}$ (3.0, 3.0\s) & $<-1.80$ (1.8, 2.4\s) & $<-1.74$ (1.9, 2.4\s) \\
1-undecene & $-5.00^{+0.71}_{-0.69}$ (3.5, 3.1\s) & $-4.83^{+0.67}_{-0.68}$ (2.5, 2.7\s) & $<-1.53$ (1.0, 2.0\s) & $<-1.82$ (1.2, 2.2\s) \\
Isocyanic acid & $-5.70^{+0.71}_{-0.94}$ (3.5, 3.1\s) & $-5.85^{+0.72}_{-0.89}$ (2.8, 2.9\s) & $<-1.46$ ($ -0.2 $, \sd) & $<-1.25$ (0.2, 1.3\s) \\
Sodium hydride & $-2.52^{+0.85}_{-1.20}$ (3.1, 3.0\s) & $-2.29^{+0.74}_{-1.13}$ (2.8, 2.9\s) & $<-1.71$ ($ -0.3 $, \sd) & $<-1.75$ ($ -0.1 $, \sd) \\
Diisobutylene & $-4.18^{+0.76}_{-0.87}$ (3.1, 3.0\s) & $-4.17^{+0.70}_{-0.82}$ (2.7, 2.8\s) & $<-1.83$ (1.9, 2.5\s) & $-4.59^{+1.00}_{-0.91}$ (3.7, 3.2\s) \\
1-octene & $-4.97^{+0.74}_{-0.77}$ (3.0, 3.0\s) & $-5.03^{+0.80}_{-0.81}$ (2.1, 2.6\s) & $<-1.64$ (1.4, 2.3\s) & $<-1.62$ (1.7, 2.4\s) \\
Dimethyl disulfide & $-4.80^{+0.73}_{-0.68}$ (3.0, 3.0\s) & $-4.86^{+0.75}_{-0.67}$ (2.5, 2.8\s) & $<-1.33$ (1.4, 2.2\s) & $<-1.55$ (1.8, 2.4\s) \\
Germane & $-3.43^{+1.10}_{-1.19}$ (2.9, 2.9\s) & $-3.24^{+1.21}_{-1.18}$ (3.9, 3.3\s) & $<-1.53$ (0.1, 1.2\s) & $<-1.51$ (0.6, 1.7\s) \\
1,3-butadiene & $-3.75^{+1.03}_{-1.02}$ (2.8, 2.9\s) & $<-2.08$ (1.9, 2.5\s) & $<-1.12$ (0.1, 1.2\s) & $<-1.34$ (1.5, 2.3\s) \\
Tetrahydrothiophene & $-4.38^{+0.74}_{-0.84}$ (2.7, 2.8\s) & $<-2.99$ (1.8, 2.5\s) & $<-1.20$ ($ -0.1 $, \sd) & $<-1.27$ (0.4, 1.5\s) \\
N,n-diethylaniline & $-5.40^{+0.74}_{-0.93}$ (2.4, 2.7\s) & $<4.15$ (1.7, 2.4\s) & $<-1.12$ ($ -0.3 $, \sd) & $<-1.57$ (0.0, 1.1) \\
Butyl isocyanate & $-6.55^{+0.78}_{-0.84}$ (2.4, 2.7\s) & $-6.34^{+0.96}_{-0.94}$ (3.3, 3.1\s) & $<-1.37$ (0.3, 1.5\s) & $<-1.68$ (1.3, 2.2\s) \\ 
Allyl alcohol & $-4.81^{+0.74}_{-0.68}$ (2.3, 2.7\s) & $-4.73^{+0.84}_{-0.79}$ (2.3, 2.7\s) & $<-1.25$ ($ -0.1 $, \sd) & $<-1.22$ (0.1, 1.2\s) \\
Isoprene & $-4.84^{+0.75}_{-0.68}$ (2.3, 2.7\s) & $<-3.20$ (1.7, 2.4\s) & $<-1.28$ (0.5, 1.6\s) & $-3.33^{+0.81}_{-1.38}$ (2.5, 2.8\s) \\
Toluene diisocyanate & $-7.24^{+0.86}_{-0.99}$ (2.2, 2.6\s) & $<-5.62$ (1.0, 2.0\s) & $<-1.33$ ($ -0.3 $, \sd) & $<-1.51$ (0.3, 1.4\s) \\
Dimethyl sulfide & $-4.51^{+0.74}_{-0.79}$ (2.2, 2.6\s) & $<-3.17$ (1.5, 2.3\s) & $<-1.37$ (1.4, 2.3\s) & $-3.67^{+1.04}_{-1.21}$ (2.7, 2.8\s) \\
1-decene & $-5.29^{+0.81}_{-0.84}$ (2.1, 2.6\s) & $<-3.74$ (1.5, 2.3\s) & $<-1.58$ (1.1, 2.1\s) & $<-1.86$ (1.6, 2.3\s) \\
Pentyl nitrate & $-5.11^{+0.76}_{-0.79}$ (2.1, 2.6\s) & $<-3.79$ (1.3, 2.2\s) & $<-2.11$ (1.2, 2.2\s) & $<-1.56$ (0.7, 1.8\s) \\
Thiophosgene & $-4.17^{+0.79}_{-0.95}$ (2.0, 2.6\s) & $-4.26^{+0.82}_{-0.86}$ (2.1, 2.6\s) & $<-1.73$ ($ -0.8 $, \sd) & $<-4.38$ ($ -0.7 $, \sd) \\
Diethyl sulfide & $-4.87^{+0.83}_{-0.80}$ (2.0, 2.6\s) & $<-3.32$ (1.6, 2.3\s) & $-3.71^{+1.07}_{-1.51}$ (2.2, 2.6\s) & $-3.58^{+0.81}_{-1.18}$ (3.1, 3.0\s) \\

\hline
\end{tabular}
\caption{Retrieved $\log_{10}$ mixing ratios and model preference for different chemical species, for retrievals with the MIRI and NIR data with \texttt{POSEIDON}. The species shown have moderate evidence ($\ln \mathrm{B}\geq2.5\pm0.5$, or $\geq2.7\pm0.2\sigma$) in the NIR one-offset retrievals. We perform the NIR retrievals using two different offset cases, as described in Section~\ref{sec:retrieval}, and we conduct retrievals on MIRI data reduced via two independent pipelines, \texttt{JExoRES} and \texttt{JexoPipe}. These retrievals were performed using a baseline model of CH$_4$, CO$_2$, CO, and H$_2$O, in addition to chemical species X. The preference for species X is established by comparing the Bayesian evidence of this model to the baseline model. We give the 95$\%$ upper limit for cases where the model preference is $\ln B <2.0 $, and we omit the evidence when this is $<0.0$. Typical uncertainties on the values of $\ln B$ are $\sim \pm 0.5$. Bayesian evidence ($\ln Z$) values for the reference model are: 17796.23 for NIR with one offset; 17794.76 for NIR with two offsets; 208.42 for the MIRI \texttt{JExoRES} data; and 199.69 for MIRI \texttt{JexoPipe}. }
\label{tab:poseidonevidence}
\end{table*}

\subsection{Atmospheric Constraints} \label{sec:finalretrieval}

In order to derive reliable atmospheric constraints from the data we now consider a model that includes the prominent CNO molecules as well as a set of additional species that represent the excess absorption observed in the data. Having identified 2 molecules that show at least moderate ($\ln B=2.5\pm0.5$, or $\geq2.7\pm0.2\sigma$) evidence across all the retrievals in the NIR and MIR, we now perform a larger retrieval including these 2 molecules in addition to the 6 key CNO molecules, CH$_4$, CO$_2$, CO, H$_2$O, NH$_3$ and HCN. We refer to this model as ``6+2X''. For this purpose, we conduct atmospheric retrievals with the only the NIR data, given their much higher signal-to-noise compared to the MIRI data. 
Figure~\ref{fig:spectrum} shows the retrieved fit to the NIR transmission spectrum for this model set-up, with the corresponding posterior distributions shown in Figure~\ref{fig:posteriors} for this and other cases. In order to assess the robustness of the derived constraints we also consider a ``4+2X'' model which excludes NH$_3$ and HCN for which strong upper-limits were placed in the larger retrievals. In Figure~\ref{fig:posteriors}, we also show results from two additional canonical retrievals with only one of the top two molecules included, i.e. isobutylene or 1-pentene. 

\subsubsection{Molecular Abundances} \label{sec:res:molabundances}

The retrieved abundances and 95$\%$ upper limits for the six CNO species are shown in Table~\ref{tab:abundances}. As expected, we find evidence for CH$_4$, which is retrieved with volume mixing ratio $\log$(X$_{\text{CH}_4}$) = $-2.34_{-0.80}^{+0.71}$ for the one-offset retrieval in the ``6+2X'' setup. This is consistent with the two offset case for the same setup, for which we find a mixing ratio of $\log$(X$_{\text{CH}_4}$) = $-2.44_{-0.76}^{+0.71}$, and with the two ``4+2X'' cases considered. The 95$\%$ upper limits for the remaining molecules are also shown in Table~\ref{tab:abundances}. As in the initial retrieval, for CO$_2$ and CO, the abundances are relatively unconstrained. In the one offset case with a ``6+2X'' setup, we find upper limits of $<$$-2.56$ and $<$$-1.72$ respectively, which potentially allow for high abundances. Similarly, for H$_2$O we find a poorly constrained upper limit of $<$$-1.78$ in the ``6+2X'' one offset case. The non-detections of NH$_3$ and HCN are relatively unchanged, with $95\%$ upper limits of $<$$-4.44$ and $<$$-2.89$ respectively in the one offset case. The constraints are consistent across all the four retrievals involving the 2X species. On the other hand, the CH$_4$ abundance is systematically higher by up to $\sim$ 1 dex in the M23 set-up where these molecules are excluded, while the constraints on the other species are relatively consistent.

\begin{figure*}
	\includegraphics[width=\textwidth]{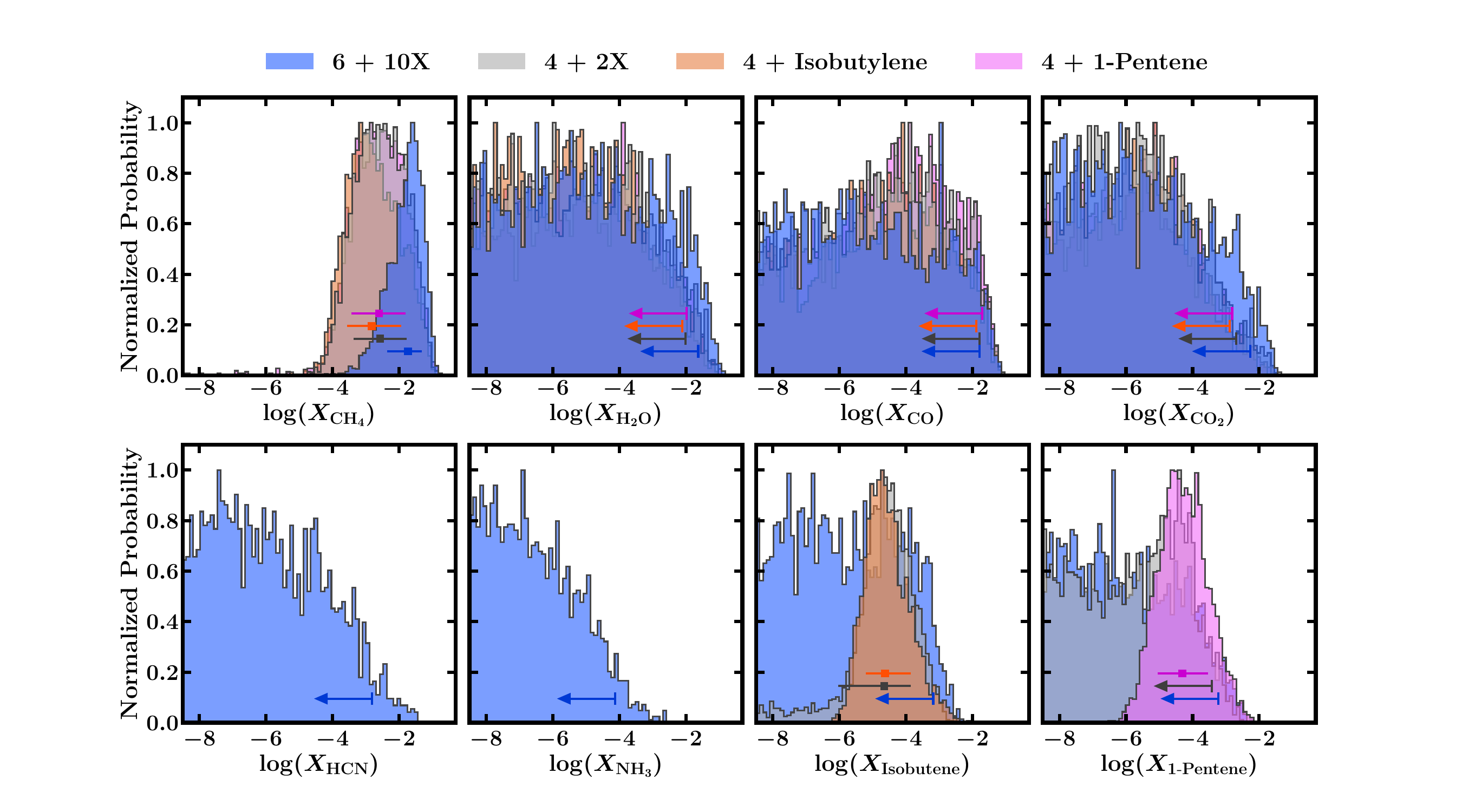}
 \caption{The posterior probability distributions retrieved for the mixing ratios of key molecules for the retrieval cases described in Section~\ref{sec:finalretrieval}. Horizontal error bars denote the median and corresponding 1$\sigma$ interval for the molecules with significant evidence while the arrows denote the 2$\sigma$ upper limits. ``6+4X'' is the model described in Section~\ref{sec:finalretrieval}, consisting of the 6 key CNO molecules (CH$_4$, CO$_2$, CO, H$_2$O, NH$_3$, and HCN) and the 4 chemical species X that exhibit moderate preference across the retrieval cases in this work (1-pentene, isobutylene, diisobutylene, and diethyl sulfide). ``4+4X'' is the baseline model (CH$_4$, CO$_2$, CO, H$_2$O) used in Section~\ref{sec:tracespecies} along with the 4 moderate-preference chemical species. The remaining models (``4+X'') show the baseline model with one of isobutylene and 1-pentene. All models consider one offset between the NIRISS and NIRSpec data. Evidence for CH$_4$ is found at a significance of $\ln B \sim 3.1$ ($\sim$3$\sigma$). Furthermore, evidence for excess absorption beyond the 6 expected CNO species is also present in the near infrared. As discussed in Section \ref{sec:tracespecies}, this may be due to a number of degenerate species. These include isobutylene and 1-pentene, which we include here. As shown, while each of these species presents a distinct peak in the respective ``4+X'' model, only one (isobutylene, in this case) does when multiple complex species are considered at once. The abundance estimates and detection significances are shown in Table~\ref{tab:abundances}.
 }
\label{fig:posteriors}
\end{figure*}

We additionally performed retrievals with a broader set of possible trace species, shown in Table~\ref{tab:abundances} as ``6+10X''. The 10 molecules used are those with the highest evidence in the 1-offset NIR retrievals, i.e. the first 10 molecules in Table~\ref{tab:poseidonevidence}. As expected and as demonstrated in Figure~\ref{fig:posteriors}, the evidence for isobutylene and 1-pentene is lower in the 6+10X case than the 6+2X case, due to the larger number of parameters and the degeneracy between the possible trace species.

\begin{table*}
\centering

\begin{tabular}{lccccccccc}
\hline \hline
Setup & Offsets & CH$_4$ & CO$_2$ & CO & H$_2$O & NH$_3$ & HCN & $T_{10\mathrm{mbar}}$ (K) & ln Z \\
\hline
6 + 10X & 1 offset &  $-1.73_{-0.64}^{+0.41}$ (3.3, 3.1$\sigma$) & $<-2.28$ & $<-1.79$ & $<-1.63$ & $<-4.13$ & $<-2.81$ & $349.9_{-67.9}^{+72.2}$ & $17801.33$\\
6 + 10X & 2 offsets & $-1.70_{-0.62}^{+0.41}$ (3.2, 3.0$\sigma$) & $<-2.18$ & $<-1.79$  & $<-1.68$ & $<-4.15$ & $<-2.62$ & $339.6_{-67.9}^{+75.0}$ & $17799.06$\\
6 + 2X & 1 offset &  $-2.34_{-0.80}^{+0.71}$ (3.6, 3.2$\sigma$) & $<-2.56$ & $<-1.72$ & $<-1.78$ & $<-4.44$ & $<-2.89$ & $310.6_{-58.3}^{+78.5}$ & $17800.85$\\
6 + 2X & 2 offsets & $-2.44_{-0.76}^{+0.71}$ (3.5, 3.1$\sigma$) & $<-2.76$ & $<-1.77$  & $<-1.90$ & $<-4.80$ & $<-3.65$ & $309.9_{-58.1}^{+77.5}$ & $17798.14$\\
4 + 2X & 1 offset & $-2.57_{-0.80}^{+0.79}$ (4.0, 3.3$\sigma$) & $<-2.68$ & $<-1.79$ & $<-2.01$ & - & - & $300.7^{+89.8}_{-56.1}$ & $17801.43$ \\
4 + 2X & 2 offsets &  $ - 2.61^{+0.81}_{-0.83}$ (3.7, 3.2$\sigma$) & $<-2.83$ & $<-1.70$ & $<-1.81$ & - & - & $320.1^{+86.8}_{-62.8}$ & $17799.48$ \\

M23 & 1 offset &  $-1.57^{+0.36}_{-0.48}$ (8.8, 4.6$\sigma$) & $<-2.31$ & $<-1.67$ & $<-1.50$ & $<-4.22$ & $<-3.06$ & $350.0^{+68.7}_{-58.8}$ & $17797.38$ \\ 
M23 & 2 offsets & $-1.61^{+0.43}_{-0.55}$ (8.6, 4.5$\sigma$) & $<-2.32$ & $<-1.57$&$<-1.49$ & $<-4.36$ & $<-3.17$ & $332.4^{+73.7}_{-67.2}$ & $17795.26$ \\
\hline
\end{tabular}
\caption{Molecular abundances retrieved for prominent molecules in the atmosphere of TOI-732~c and their detection significances. These are shown for each of the cases described in Section~\ref{sec:finalretrieval} for different combinations of offsets between the NIR instruments. We show the molecular abundances as $\log_{10}$ of the volume mixing ratios. The median and 1$\sigma$ estimates are given for CH$_4$, which shows strong to moderate evidence. For the remaining molecules, the 95\% upper-limits are given. For CH$_4$, the detection significances $>1\sigma$ are shown in brackets. Each detection significance has a nominal $\sim$0.1$\sigma$ uncertainty which results from the uncertainty on the Bayesian evidence as estimated during nested sampling. ``6+2X'' is the model described in Section~\ref{sec:finalretrieval}, consisting of the 6 key CNO molecules (CH$_4$, CO$_2$, CO, H$_2$O, NH$_3$, and HCN) and the 2 chemical species X that exhibit moderate preference across the retrieval cases in this work (1-pentene and isobutylene). ``4+2X'' is the baseline model with four prominent molecules (CH$_4$, CO$_2$, CO, H$_2$O) used in Section~\ref{sec:tracespecies} along with the 2X chemical species. M23 is a model that includes the same species as the \citet{Madhusudhan2023b} canonical case, i.e., the six CNO species, and five predicted possible biosignature species (DMS, CS$_2$, CH$_3$Cl, OCS, N$_2$O).}
\label{tab:abundances}
\end{table*}

\subsubsection{Clouds/Hazes}

We show the constraints for the cloud/haze parameters in Table~\ref{tab:physical_properties}. The transmission spectrum provides nominal constraints on the clouds/haze parameters at the terminator. For the preferred one-offset case, a model with clouds/hazes present is preferred over a cloud/haze-free model by $\ln B=$ 5.66 (4$\sigma$). We find evidence for clouds/hazes, with the cloud-top pressure for gray clouds at $\log(P_c/\mathrm{bar}) \sim -2.5$. The scattering slope $\gamma$ is not well constrained, however the Rayleigh enhancement factor $a$ is found to be significantly larger than expected for H$_2$ Rayleigh scattering alone. We find a clouds/haze coverage fraction at the terminator of $\sim$$0.6\pm0.1$. More detailed modeling of clouds/hazes may refine the corresponding constraints on the cloud properties. 

\subsubsection{Temperature Structure}

Nominal constraints can additionally be placed on the photospheric temperature at the terminator. In the one offset 6+2X case, we find a temperature of $310.6_{-58.3}^{+78.5}$~K at $10^{-2}$~bar, as shown in Table~\ref{tab:abundances}. The derived photospheric temperature is consistent within the 1$\sigma$ uncertainties across the cases considered. Given the retrieved photospheric temperature, this could suggest the presence of a cold trap, such that H$_2$O condenses out. However, this is not strictly required by the relatively unconstrained H$_2$O abundance. 

Similar to the suggestion for K2-18~b \citep{Madhusudhan2023b}, the presence of clouds/hazes in the atmosphere of TOI-732~c could allow for temperate conditions at the surface. Emission spectroscopy would be required to better constrain the atmospheric temperature structure and Bond albedo, providing further insight into the potential habitability of TOI-732~c. Additional transmission spectra in this wavelength range and in the optical could also place improved constraints on the presence and properties of clouds/hazes in the atmosphere.

\begin{table*}
\centering

\begin{tabular}{lcccccccc}
\hline \hline
Setup & Offsets &  $\phi$ & $\log a$ & $\gamma$ & $\log (P_{\mathrm{c}}\mathrm{/bar})$ & $\log (P_{\mathrm{ref}}\mathrm{/bar})$ & OS1/ppm & OS2/ppm \\
\hline 

6 + 10X & 1 offset & $0.64^{+0.09}_{-0.10}$ & $ 8.71^{+0.84}_{-1.24}$ & $-9.15^{+1.91}_{-1.57}$ & $ - 2.21^{+1.29}_{-1.31}$ & $-4.54^{+0.51}_{-0.36}$ & $ -92.15_{-24.66}^{+23.38}$ & - \\
6 + 10X & 2 offsets &  $ 0.63^{+0.10}_{-0.10}$ & $ 8.69^{+0.87}_{-1.40}$ & $- 9.13^{+1.98}_{-1.66}$ & $- 2.08^{+1.33}_{-1.25}$ & $- 4.49^{+0.53}_{-0.38}$ & $-92.94_{-25.01}^{+23.49}$ & $-88.70_{-27.82}^{+25.14}$ \\
6 + 2X & 1 offset & $0.56^{+0.10}_{-0.09}$ & $8.49^{+1.00}_{-1.41}$ & $ - 9.20^{+2.01}_{-1.84} $ & $ - 2.93^{+1.52}_{-0.94} $ & $ - 3.95^{+0.57}_{-0.53} $ & $-112.38_{-26.53}^{+24.95}$ & - \\
6 + 2X & 2 offsets &  $ 0.55^{+0.10}_{-0.09} $ & $ 8.52^{+0.97}_{-1.54} $ & $ - 9.20^{+1.98}_{-1.93} $ &$ - 2.88^{+1.34}_{-0.95} $ & $ - 3.84^{+0.54}_{-0.54} $ & $-112.00_{-26.90}^{+26.94}$ & $-116.54_{-28.36}^{+28.64}$ \\
4 + 2X & 1 offset &  $ 0.55^{+0.10}_{-0.10} $ & $ 8.55^{+0.98}_{-1.49} $ & $ - 9.36^{+2.09}_{-1.89} $& $ - 2.78^{+1.52}_{-1.01} $ & $ - 3.78^{+0.57}_{-0.56} $ & $-113.03_{-25.36}^{+24.92}$ & - \\
4 + 2X & 2 offsets & $ 0.55^{+0.10}_{-0.10} $ & $ 8.55^{+0.99}_{-1.49} $ & $ - 9.20^{+2.00}_{-1.84} $ & $ - 2.86^{+1.41}_{-0.97} $ & $ - 3.81^{+0.62}_{-0.54} $ & $-113.18_{-26.92}^{+27.17}$ & $-118.63_{-29.44}^{+29.00}$ \\

M23 & 1 offset &  $ 0.60^{+0.11}_{-0.11}$ & $ 8.35^{+1.14}_{-1.62}$ & $-8.85^{+2.32}_{-1.96}$ & $-3.29^{+1.05}_{-0.73}$ & $-4.36^{+0.49}_{-0.41}$ & $ -125.80_{-23.06}^{+23.43}$ & - \\
M23 & 2 offsets &  $ 0.58^{+0.11}_{-0.12}$ & $ 8.32^{+1.14}_{-1.74}$ & -$ 8.97^{+2.21}_{-2.24}$ & $- 3.14^{+1.31}_{-0.84}$ & $- 4.24^{+0.52}_{-0.45}$ & $-126.34_{-23.66}^{+21.79}$ & $-117.79_{-27.57}^{+28.00}$ \\
\hline
\end{tabular}
\caption{The retrieved properties of the atmosphere of TOI-732~c, including the temperature, cloud/haze properties and reference pressure. The retrieved dataset offsets with respect to NIRISS are also shown, for the models described in Section~\ref{sec:finalretrieval}. ``6+2X'' is the model described in Section~\ref{sec:finalretrieval}, consisting of the 6 key CNO molecules (CH$_4$, CO$_2$, CO, H$_2$O, NH$_3$, and HCN) and the 2 chemical species X that exhibit moderate preference across the retrieval cases in this work (1-pentene and isobutylene). ``4+2X'' is the baseline model (CH$_4$, CO$_2$, CO, H$_2$O) used in Section~\ref{sec:tracespecies} along with the 2 moderate-preference chemical species. M23 is a model that includes the same species as the \citet{Madhusudhan2023b} canonical case, i.e., the six CNO species, and five predicted possible biosignature species (DMS, CS$_2$, CH$_3$Cl, OCS, N$_2$O).  }
\label{tab:physical_properties}
\end{table*}

\subsubsection{Stellar Heterogeneities}

We also perform retrievals that include stellar heterogeneities which could affect the transmission spectrum. For this purpose, we investigate the evidence for a model including stellar heterogeneities relative to the baseline model with the four prominent molecules. This is the most conservative case, aimed to investigate if stellar heterogeneities could account for the excess absorption required by the data. We find no significant evidence for unocculted stellar heterogeneities affecting the spectrum. 

\subsubsection{Detection Significances}

We determine the detection significances for the molecules in each of the above retrievals, as shown in Table~\ref{tab:abundances}. CH$_4$ is found to have the strongest detection significance across the prominent CNO molecules and across all retrieval cases. For the 4+2X one offset retrieval we detect CH$_4$ with evidence $\ln B =4.0$ (3.3$\sigma$). The one offset case is preferred over the two offset case by $\ln B=1.95$. The lower significance of CH$_4$ in these retrievals, compared to the $>$4$\sigma$ evidence using the \citet{Madhusudhan2023b} setup, is due to the fact that in the absence of CH$_4$ much of its absorption in the NIR can be compensated for by degenerate features from one or more complex molecules in the model, represented by 2X. On the other hand, the detection significance for the 2X molecules together is $\ln B=5.2$ ($\sim$3.7$\sigma$), for the 4+2X one-offset set-up. The significance of each molecule cannot be resolved individually, due to the strong degeneracy between them. However, the Bayesian evidences of retrievals with the 6+2X and 4+2X models are comparable with each other as well as with a 4+1X model, where the 1X is either isobutene or 1-pentene. Therefore, there is significant evidence for excess absorption in the NIR data at $\ln B\sim5.2$ ($\sim$$3.7\sigma$) significance in the preferred one-offset case. We also note that the cross sections for these complex species may be incomplete, which could further affect the constraints on CH$_4$ in future works.

The overall significance for the presence of at least one of these 2 additional molecules, compared to the baseline model of 6 molecules (CH$_4$, CO$_2$, CO, H$_2$O, NH$_3$ and HCN) is $\ln B=5.27$ ($\sim$3.7$\sigma$) in the NIR with one offset. The significance of each molecule cannot be resolved individually, due to the strong degeneracy. These species are also degenerate with CH$_4$. This is evident in the difference in CH$_4$ significance between the \citet{Madhusudhan2023b} set-up ($\ln B =8.8$, 4.6$\sigma$) and 6+2X set-up ($\ln B =3.7$, 3.2$\sigma$), as shown in Table~\ref{tab:abundances}, due to the presence of different absorbing species.

We find evidence of excess absorption in the transmission spectrum of TOI-732~c at $\ln B\sim5.2$ ($\sim$$3.7\sigma$), and have identified species that result in moderate evidence across all our retrieval cases. However, our analysis is not exhaustive of all possible chemical species, and the current quality of the data means it is not possible to robustly determine the cause of the excess absorption. Additional observations will be required to increase the precision and allow refined constraints for the potential molecules. This will be discussed further in Section~\ref{sec:discussion}.

\section{Summary and Discussion} \label{sec:discussion}

We report the first transmission spectrum of the temperate sub-Neptune TOI-732~c observed with JWST. One transit was observed with each of NIRISS SOSS, NIRSpec G395H, and MIRI LRS, resulting in a transmission spectrum spanning the 0.9-12 $\mu$m range. We find moderate to strong evidence ($\ln B$ = 3.2-8.8, 3.0-4.6$\sigma$) for CH$_4$ and place upper limits on H$_2$O, CO$_2$ and CO, which are relatively unconstrained but with 95$\%$ upper limits allowing high abundances. We additionally report non-detections of NH$_3$ and HCN with 95$\%$ upper limits on their log mixing ratios at $\lesssim$$-4.5$ and $\lesssim$$-3.5$, respectively. We also find evidence for excess absorption in the data beyond that allowed by these prominent CNO molecules. Therefore, we conduct an extensive exploration of 250 chemical species to investigate potential contributors to the excess absorption. 

We find evidence for additional absorption due to one or more complex molecules, including higher order hydrocarbons and/or sulfur-bearing species. Across the 250 molecules explored, we identify the species that exhibit at least moderate evidence ($\ln B\geq2.5\pm0.5$) while considering the different datasets. Here, we consider the evidence in favor of the addition of each chemical species X to a baseline model comprising prominent molecules which can be abundant in the atmosphere as noted above: CH$_4$, CO$_2$, CO and H$_2$O. We find 29 molecules with evidence $\geq2.5\sigma$ in the NIR data, with the highest evidence for a molecule found at $\ln B=5.9$ (3.9$\sigma$). We further find only 2 of these molecules that also show moderate evidence in the MIR data from the two independent reductions \texttt{JExoRES} and \texttt{JexoPipe}. These two molecules are isobutylene and 1-pentene. Several other molecules also show moderate evidence in at least one of the two wavelength ranges. In particular, six additional molecules show $\ln B \geq 2.0$ ($\geq2.5\sigma$) evidence in the NIR data and one of the two MIR reductions. These are propylene (C$_3$H$_6$), 2-methyl-1-butene (C$_5$H$_{10}$),   diisobutylene (diisobutene or 2,4,4-trimethyl-1-pentene, C$_8$H$_{16}$), DMS (CH$_3$SCH$_3$), isoprene (C$_5$H$_8$), and diethyl sulfide (C$_4$H$_{10}$S).  In all cases, the abundance of the relevant molecule is typically retrieved at a log mixing ratio between -6 and -4.

Overall, we find evidence for excess absorption due to additional chemical species at $\ln B\sim5.2$ ($\sim$$3.7\sigma$), compared to a baseline model comprising only the six prominent CNO molecules typically considered in atmospheric retrievals of sub-Neptunes. This is the evidence corresponding to the combination of the two complex molecules noted above with significant evidence across the data sets, while each molecule individually is favored between $\ln B= 2.0-5.9$ (2.5-3.9$\sigma$) relative to the baseline model. Future observations are required to resolve the degeneracy between and provide improved constraints for possible trace species in the atmosphere of TOI-732~c. 

We report nominal constraints on the other atmospheric properties at the day-night terminator of the planet. We retrieve a slant photospheric temperature of $\sim$310 $\pm$60 K at 10 mbar, which is consistent with the equilibrium temperature of the planet. The retrieved temperature range also allows for a cold trap at the photosphere where H$_2$O could condense out. However, the relatively unconstrained H$_2$O abundance does not require the presence of a cold trap. We also find significant evidence ($\ln B\sim5.7$) for the presence of clouds/hazes at the terminator with the inhomogeneous cloud parametrization adopted in the present framework. However, more detailed cloud modeling may refine the corresponding constraints on the cloud properties. 

The bulk properties of TOI-732~c, as for other temperate sub-Neptunes of similar density, can be explained by a number of degenerate interior compositions. These include a mini-Neptune, a gas dwarf, and a hycean world \citep{Madhusudhan2021, Rigby2024}. In what follows, we discuss the implications of the possible trace species in the atmosphere of TOI-732~c, the possible nature of this planet based on our chemical abundance constraints, and the prospects for future observation and characterization of TOI-732~c and similar planets. 

\subsection{Implications of Trace Molecules} \label{sec:disc:tracegases}

A central finding in this work is the requirement for one or more complex molecules in the atmosphere of TOI-732 c to explain the data. As discussed above, across the 250 molecules explored in this work, only 2-8 molecules show indications of moderate evidence across the NIR and MIR data. These include isobutylene (isobutene), 1-pentene, diisobutylene, diethyl sulfide, propylene, 2-methyl-1-butene, DMS and isoprene, with one or more of them retrieved at abundances between 1-100 ppmv. Currently, the contributions from these species are degenerate thereby precluding a definitive detection of any one of them, with the possibility that hitherto unconsidered species may also be potential candidates. All these species are known to be trace gases on Earth, with abundances below 10 ppb, originating primarily from biogenic or anthropogenic sources. While DMS and isoprene have been predicted to be promising biosignatures in H$_2$-rich atmospheres \citep[e.g.][]{domagal-goldman2011, seager2013b, Madhusudhan2021, Zhan2021}, the remaining molecules have not been considered in exoplanetary atmospheres nor found in significant quantities in the solar system. We note that propylene has been detected elsewhere in the solar system at low abundance, e.g. in Titan's atmosphere at $\sim$10 ppbv \citep{Nixon2013}.

The consideration of any of the above species in TOI-732~c requires a careful assessment of their physical plausibility given the environmental context. Hydrocarbons have been predicted and observed in the atmospheres of solar system bodies, including Titan \citep[e.g.][]{Vuitton2019,Titan_Nixon2024} and the giant planets \citep[e.g.][]{Moses2005, Fouchet2009, Orton2014, Moses2020, Fletcher2023}, as photochemical products of methane. Whether hydrocarbons such as isobutylene, diisobutylene and 1-pentene can be abiotically produced and maintained in such high quantities as inferred from the retrievals is unclear, and none of these particular molecules have been detected to date in any planetary atmosphere other than Earth. Furthermore, we note that aside from methane, we find no evidence for simpler hydrocarbons, such as ethane, propane, ethylene, and acetylene. Lower-order hydrocarbons would typically be expected to be photochemically produced in larger abundances than these higher-order, more complex hydrocarbons, as evidenced in solar-system atmospheres \citep[e.g.][]{Moses2005,Titan_Nixon2024}. Ion chemistry could potentially enhance the abundances of complex hydrocarbons, as on Titan \citep[e.g.][]{Vuitton2019}, although lighter hydrocarbons are still expected to dominate. Therefore, future work is required to investigate the plausibility of producing any of the above complex molecules in TOI-732~c, whether in a deep or shallow H$_2$-rich atmosphere, through abiotic or biotic mechanisms. 

In this study we conduct a wide exploration of 250 possible trace species. Our exploration of trace species in Section~\ref{sec:tracespecies} also includes lower-order hydrocarbons. We do not find strong constraints on these molecules, but our current constraints could allow for high abundances -- for example, in the NIR we obtain an upper limit of $<$$-1.60$ for ethane. However, we also note that the chemical species explored in this study do not comprise a comprehensive list of all possible molecules, and that combinations of species have not been explored. 

A relevant question is whether any of these molecules can be indicators of biological activity on the planet. The planet has been proposed as a candidate hycean world \citep{Madhusudhan2021}. However, the possibility of an ocean requires an adequate albedo from clouds/hazes \citep{Madhusudhan2021, Leconte2024}. While we do find significant evidence for clouds/hazes at the terminator, it remains to be seen if the required day side albedo for habitability is feasible for this planet. If that is the case, then it is possible that at least some of the candidate complex molecules may be contributed by biological activity. For example, DMS and isoprene can both reach high abundances, up to 100 ppm, in H$_2$-rich atmospheres with underlying biological activity \citep{Tsai2021,Zhan2021}. Some of the other molecules, such as isobutene, are also known to be produced in small quantities by some microbial life on Earth \citep{1987isobutene}. However, it is unclear if biotic ocean sources could produce such large quantities of the complex hydrocarbons in the atmosphere of this planet.

\subsection{Possible Interior Scenarios} \label{sec:disc:interior}

Our inferred atmospheric composition of TOI-732~c provides nominal constraints on its interior composition. The bulk properties of TOI-732~c are compatible with degenerate interior compositions spanning hycean world, mini-Neptune and gas dwarf scenarios, similar to K2-18~b \citep{Madhusudhan2021, Rigby2024, Rigby2024b}, all with an H$_2$-rich atmosphere. While the hycean scenario requires a shallow atmosphere, the other two scenarios involve a deep atmosphere. 

The observable signatures of a surface ocean beneath a thin H$_2$-rich atmosphere were outlined in \citet{Madhusudhan2023a}, based on photochemical modelling studies \citep{Yu2021,Hu2021,Tsai2021}. These studies suggest that a high detected abundance of CH$_4$ combined with the nondetection of NH$_3$ could be one indicator of a hycean ocean. The presence of substantial CO$_2$ in the atmosphere, with CO$_2$$>$CO have also been suggested to be indicative of possible hycean conditions when H$_2$O is not detected \citep{Hu2021,Madhusudhan2023a}. The lack of strong evidence for H$_2$O in the photosphere at the terminator of TOI-732~c is consistent with the presence of a tropospheric cold trap \citep{Madhusudhan2023a}, such that the H$_2$O is condensed out at pressures higher than those probed by transmission spectroscopy. We note the presence of a liquid water ocean in the hycean scenario would require an adequate albedo \citep{Madhusudhan2021, Piette2020,  Leconte2024}. 

In the case of TOI-732~c, neither CO$_2$ nor CO are well constrained by the observations, with their upper limits permitting high abundances. As pointed out by \citet{Madhusudhan2023a}, the absence of CO$_2$ is relatively uninformative for the inference of potential hycean conditions, as it could be due to, for example, CO$_2$ dissolution in the ocean \citep{Hu2021}. The current atmospheric constraints for TOI-732~c therefore do not rule out the possibility of a surface ocean atop an icy mantle \citep[e.g.][]{Rigby2024}. Alternatively, as suggested for K2-18~b \citep{Luu2024}, the surface could be supercritical -- the effect on the atmospheric chemistry in this case of supercritical H$_2$O remains relatively unexplored. At the same time, the mini-Neptune and gas dwarf scenarios, requiring deep H$_2$-rich atmospheres, may also be feasible. While the lack of NH$_3$ may be consistent with the hycean scenario, its relevance for the gas dwarf scenario depends on the feasibility of a surface magma ocean \citep[e.g.][]{Rigby2024b}. The lack of strong upper limits for both CO and CO$_2$ mean that these observations are inconclusive with regards to the deep vs shallow atmosphere scenarios \citep{Hu2021,Wogan2024,Cooke2024, Rigby2024b}.

It is possible that the presence of abundant complex hydrocarbons may be indicative of a shallow H$_2$-rich atmosphere over an ocean. In such an atmosphere the photochemical products of CNO chemistry are prevented from recycling back to the more stable thermochemical-equilibrium parent species CH$_4$, NH$_3$, and H$_2$O at depth. This can result in low NH$_3$ as well as potentially large abundances of heavy complex hydrocarbons \citep[e.g.][]{Madhusudhan2023a,Huang2024}, accumulated over time. Based on the bulk density constraints, a shallow atmosphere in TOI-732~c requires a large mass fraction of H$_2$O in the interior, resulting in an ocean surface, e.g. a hycean world scenario, or a supercritical ocean. Further research is also needed to investigate the consequences of chemistry and atmospheric mixing in a deep-atmosphere scenario. The sharp change in density and mean molecular mass caused by a heavy species, such as water, condensing in an H$_2$-dominated atmosphere could act as a barrier to convective mixing \citep[e.g.,][]{Leconte2024}. This may act to prevent or extend the time scales for photochemical products being transported to higher-temperature regimes where they could be recycled. Moreover, it is unclear how gas-phase chemistry and mixing proceeds in an H$_2$-H$_2$O atmosphere when temperature and pressures exceed the critical point of H$_2$O \citep{Gupta2025}. We highlight the need for further study in these areas.

\subsection{Future Directions}
\label{sec:disc:future}

Alongside K2-18~b and TOI-270~d \citep{Madhusudhan2023b,Holmberg2024,Benneke2024,Madhu2025}, TOI-732~c is the third temperate sub-Neptune observed with JWST to date, with each set of observations revealing detections of carbon-bearing molecules in the planetary atmosphere. Future JWST observations of these planets will add greater precision, potentially providing improved constraints on the atmospheric properties and composition. To robustly establish the nature of sub-Neptune exoplanets, atmospheric data must be linked to atmospheric and interior models to break internal structure degeneracies \citep[e.g.][]{Rigby2024b,Nixon2024}. The high-precision atmospheric data achievable with JWST motivates the need for further theoretical and experimental studies to robustly inform characterization efforts, including atmospheric chemistry and surface-atmosphere interactions.

Additional observations of TOI-732~c are required to increase the precision sufficiently to resolve the degeneracy between complex molecules potentially contributing to the excess absorption. Our exploration of possible trace species in both the MIRI and NIR ranges revealed 2 molecules with moderate ($\ln B \geq2.0$, or $\geq$$2.5\sigma$) evidence in both ranges, across two MIRI reduction pipelines. In future, as more stringent constraints are placed on the presence of these and/or other molecules, the physical plausibility of producing and maintaining their inferred atmospheric abundance should be considered in context. We additionally highlight the need for extensive data for the molecular cross-sections, including at wavelengths relevant to the NIRISS observations in this work, and for a broader set of molecules.

Both the startling similarities and intriguing diversity among the transmission spectra of TOI-732~c, K2-18~b \citep{Madhusudhan2023b} and TOI-270~d \citep{Holmberg2024,Benneke2024} provide exciting first insights into the nature of temperate planets in the sub-Neptune regime. These observations demonstrate the immense capability of JWST to shed light on the sub-Neptune population, with important implications for our understanding of their diversity and potential for habitability.

{\it Acknowledgements:} This work is based on observations made with the NASA/ESA/CSA James Webb Space Telescope as part of Cycle 2 GO Program 3557 (PI: N. Madhusudhan).  We thank NASA, ESA, CSA, STScI, everyone whose efforts have contributed to the JWST, and the exoplanet science community for the thriving current state of the field. NM acknowledges support from the UK Research and Innovation (UKRI) Frontier Research Grant (EP/X025179/1). N.M., F.E.R. and L.P.C. acknowledge support from UKRI STFC toward the doctoral studies of F.E.R (UKRI grant 2605554) and L.P.C (UKRI grant 2886925). N.M. thanks Tony Roman, Alaina Henry, Joseph Filippazzo and Sara Kendrew at STScI for their help with planning our JWST observations. J.M. acknowledges support from NASA Exoplanets Research Program 80NSSC22K0314 and JWST-GO-03557, which was provided by NASA through a grant from the Space Telescope Science Institute, which is operated by the Association of Universities for Research in Astronomy, Inc., under NASA contract NAS 5-03127. 

This research has made use of the NASA Exoplanet Archive, which is operated by the California Institute of Technology, under contract with the National Aeronautics and Space Administration under the Exoplanet Exploration Program. This research has made use of the NASA Astrophysics Data System and the Python packages \texttt{NUMPY}, \texttt{SCIPY}, and \texttt{MATPLOTLIB}.

This work was performed using resources provided by the Cambridge Service for Data Driven Discovery operated by the University of Cambridge Research Computing Service (\url{www.csd3.cam.ac.uk}), provided by Dell EMC and Intel using Tier-2 funding from the Engineering and Physical Sciences Research Council (capital grant EP/P020259/1), and DiRAC funding from STFC (\url{www.dirac.ac.uk}).

{\it Author Contributions:} N.M. conceived, planned and led the project. N.M. led the JWST proposal with contributions from S.S., F.E.R and J.M. F.E.R. led the writing of the manuscript with contributions from all authors. N.M., M.H. and S.S. conducted the data reduction and analyses. N.M. and L.P.C. conducted the atmospheric retrievals. F.E.R., N.M., L.P.C., and J.M. conducted the theoretical interpretation.

{\it Data Availability:} The JWST data presented in this paper were obtained from the Mikulski Archive for Space Telescopes (MAST) at the Space Telescope Science Institute. The specific observations analyzed can be accessed via \dataset[doi:10.17909/56b4-3b33]{https://doi.org/doi:10.17909/56b4-3b33}. 

The transmission spectra of TOI-732 c reported in this work are available on the Open Science Framework at \dataset[https://osf.io/wsjmu]{https://osf.io/wsjmu}.

\facilities{JWST (NIRISS, NIRSpec and MIRI)}\\

\appendix

\section{Bayesian Priors for Atmospheric Retrieval}
\label{sec:priors}

In Table~\ref{tab:retrieval_priors} we show the Bayesian prior probability distributions used in the atmospheric retrievals in this work. $\mathcal{U}(X, Y)$ represents a uniform probability distribution between values X and Y. $\mathcal{N}(\mu, \sigma^2)$ represents a normal distribution, with mean $\mu$ and variance $\sigma^2$. The uniform prior for the chemical mixing ratio is the same across all chemical species considered in this work. $\delta_\mathrm{Nirspec}$ is the offset applied to the NIRSpec data with respect to the NIRISS data, in parts-per-million. $\delta_\mathrm{NRS1}$ and $\delta_\mathrm{NRS2}$ are the equivalent offsets for each of the NIRSpec detectors NRS1 and NRS2 separately.

\begin{table*}
\centering
\begin{tabular}{l|c|l}
    Parameter & Bayesian Prior  & Description\\[0.5mm]
    \hline
    \hline
    $\mathrm{log}(X_\mathrm{X})$& $\mathcal{U}$(-12, -0.3) & Mixing ratio of each chemical species \\[0.5mm]
    $T_0 / \mathrm{K} $ & $\mathcal{U}$(100, 500)& Reference temperature at 10$^{-6}$~bar \\[0.5mm]
    $\alpha_1 / \mathrm{K}^{-\frac{1}{2}}$  & $\mathcal{U}$(0.02, 2.00) & P-T profile curvature \\[0.5mm]
    $\alpha_2/ \mathrm{K}^{-\frac{1}{2}}$& $\mathcal{U}$(0.02, 2.00)  & P-T profile curvature \\[0.5mm]

    $\mathrm{log}(P_1/\mathrm{bar})$   & $\mathcal{U}$(-6, 0) & P-T profile region limit\\[0.5mm]
    $\mathrm{log}(P_2/\mathrm{bar})$  & $\mathcal{U}$(-6, 0) & P-T profile region limit \\[0.5mm]
    $\mathrm{log}(P_3/\mathrm{bar})$   & $\mathcal{U}$(-2, 0)& P-T profile region limit \\[0.5mm]
    $\mathrm{log}(P_\mathrm{ref}/\mathrm{bar})$   & $\mathcal{U}$(-6, 0) & Reference pressure at $R_\mathrm{p}$\\[0.5mm]
    $\mathrm{log}(a)$  & $\mathcal{U}$(-4, 10)& Rayleigh enhancement factor \\[0.5mm]
    $\gamma$   & $\mathcal{U}$(-20, 2)& Scattering slope \\[0.5mm]
    $\mathrm{log}(P_\mathrm{c}/\mathrm{bar})$  & $\mathcal{U}$(-6, 0)& Cloud top pressure \\[0.5mm]
    $\phi$  & $\mathcal{U}$(0, 1)& Cloud/haze coverage fraction\\[0.5mm]
    $\delta_\mathrm{Nirspec} / \mathrm{ppm}$  & $\mathcal{U}$(-300, 300)& NIRSpec dataset offset \\[0.5mm]
    $\delta_\mathrm{NRS1} / \mathrm{ppm}$  & $\mathcal{U}$(-300, 300)& NIRSpec NRS1 dataset offset \\[0.5mm]
    $\delta_\mathrm{NRS2} / \mathrm{ppm}$ & $\mathcal{U}$(-300, 300) & NIRSpec NRS2 dataset offset \\[0.5mm]
    $T_\mathrm{phot} /$  K  & $\mathcal{N}(3358, 100^2)$& Stellar photosphere temperature\\[0.5mm]
    $T_\mathrm{het} /$  K & $\mathcal{U}(2100, 4030)$ & Stellar heterogeneity temperature \\[0.5mm]
    $f_\mathrm{het} $  & $\mathcal{U}$(0, 0.5)& Stellar Heterogeneity coverage fraction\\[0.5mm]
     \hline
\end{tabular}
\caption{The model parameters and Bayesian prior probability distributions used in the atmospheric retrievals performed with \texttt{AURA} \citep{Pinhas2018} and \texttt{POSEIDON} \citep{macdonald2017,Macdonald2024}.}
\label{tab:retrieval_priors}
\end{table*}

\section{List of Molecules Considered} \label{Appendix:Moleculeslist}

We consider the following chemical species in our exploration of possible trace species, described in Section~\ref{sec:tracespecies}:

Cyclopentene, Propylene, 2-methyl-1-butene, Isobutylene, 1-butene, 2-methyl-1-pentene, 1-pentene, 1-hexene, 1-heptene, 4-methyl-1-pentene, Diisobutylene, 1-undecene, 1-octene, Dimethyl disulfide, Dimethyl sulfide, 1,3-butadiene, 1-nonene, Isoprene, Diethyl sulfide, 1-decene, Isocyanic acid, Morpholine, (-)-beta-pinene, N,n-diethylaniline, Tetrahydrothiophene, 3-methyl-1-butene, Pentyl nitrate, Propylbenzene, Butyl isocyanate, 4-vinylcyclohexene, 1,2,3,4-tetrahydronaphthalene, Isobutyronitrile, Ethylbenzene, 2-methylaziridine, Methanethial, Cyclopentane, Cyclodecane, 2-methylpyridine, Allylamine, Thiophosgene, 2-carene, 2,4-diisocyanato-1-methylbenzene, Propionitrile, Dimethyl sulfoxide, Thiophene, Pyridine, Sulfur monoxide, Methacrylonitrile, Triethylamine, Acrylonitrile, (+)-limonene, Cycloheptene, Ethyl nitrite, Diethylamine, Myrcene, Isobutane, Methyl isocyanate, 1-ethyl-2-methylbenzene, 4-ethyltoluene, Nitrogen trifluoride, N,n-diethylformamide, 2-methylpentane, Decane, Propylene sulfide, Pentane, Pentadecane, Hexane, Nitrobenzene, Isopentane, Butane, Cyclohexane, Magnesia, Benzonitrile, Heptane, 3-methylpentane, Neopentane, Toluene, Cycloheptane, 3-ethyltoluene, 1-nitropropane, Methyl nitrite, Nicotine, Trans-2-pentene, Cyclohexene, 2,2-dimethylbutane, Octane, 2,4,4-trimethyl-2-pentene, 2-vinylpyridine, Undecane, 2-methyl-2-pentene, Butylamine, Ethylamine, Cis-4-methyl-2-pentene, Cyclooctane, 2-butene (cis and trans mixture), Propane, Tert-amylamine, Piperidine, Limonene, (-)-alpha-pinene, Ethylene, 3-methylhexane, Perchloromethyl mercaptan, 2-methyl-2-butene, Nitromethane, O-toluidine, Hexamethylphosphoramide, 1,1-dimethylhydrazine,Sec-butylbenzene, Nonane, Isopropylamine, 2,6-diethylaniline, Sec-amylamine, 1-propanethiol, Hydrazine, Tetradecane, Propyne, Barium, Methanesulfonyl chloride, Hydrogen sulfide, Sulfur dioxide, Acetylene, Hexadecane, 4-methylpyridine, Methylhydrazine, Benzene, Chloromethane, Cyclopropane, Sulfuryl fluoride, Trifluoronitrosomethane, Trimethylamine, Acetone cyanohydrin, 1-butyne, Methylamine, Sulfur hexafluoride, Aniline, 2-mercaptoethanol, 2,2,4-trimethylpentane, Nitroethane, Sulfuryl chloride, Ammonia, Ozone, 2-methyl-2-propanethiol, 2-propanethiol, Ethane, Hydroxide, Benzenethiol, Cis-2-pentene, Potassium hydroxide, Rubidium, Sodium hydroxide, Methyl isothiocyanate, Isobutyl mercaptan, Phosphine, Thiophosphoryl chloride, Ethanethiol, Methanol, Mesitylene, 1,2,3,5-tetramethylbenzene, Hydrofluoric acid, Sodium, Cyanoacetylene, Thiirane, Carbon disulfide, Styrene, Hydrogen cyanide, But-2-ynenitrile, Potassium, Vanadium oxide, Titanium, Scandium, Ethylenediamine, Lithium hydride, Cesium, Dimethyl sulfate, Acetonitrile, Manganese, Aluminum, Methyl radical, Hydrochloric acid, Silicon monoxide, Nickel, Lithium, Allene, Vanadium, Quinoline, Chromium, Tridecane, Peroxyacetyl nitrate, 2,4-dimethylpentane, Cumene, Potassium chloride, 2-nitropropane, 3-carene, Pentane-1,5-diamine, Naphthalene, Calcium, Methanethiol, Allyl isothiocyanate, Iron, 2-methylstyrene, Cyclohexanethiol, Nitrous oxide, 2,3-dimethylbutane, Thionyl fluoride, Trimethylbenzene, N,n-dimethylformamide, 1,2,3,4-tetramethylbenzene, Tert-butylbenzene, Diisopropylamine, Diethyl sulfate, Dimethylamine, Magnesium, Nitrous acid, NaH, GeH$_4$, FeH, CaH, OCS, C$_2$, TiH, CS, BeH, PO, NaO, PN, O, LaO, NO, NO$_2$, ZrO, H3+, OH+, NH, AlO, SH, ScH, PS, CH, Ba+, AlH, MgH, Ca+, SiH, Ti+, Fe+, CaO, Mg+, TiO, V+, CrH


\bibliography{ms.bib}
\bibliographystyle{aasjournal}

\end{document}